\documentclass[10pt,journal,compsoc]{IEEEtran}
\usepackage{microtype}                 
\usepackage{textcomp}                  
\usepackage{mathptmx}                  
\usepackage{times}                     
\usepackage{tabu}                      
\usepackage{booktabs}                  
\usepackage{cite}

\usepackage{paralist}
\usepackage{amssymb}
\usepackage{float}
\usepackage{placeins}
\usepackage{scrextend}
\usepackage{graphicx}
\usepackage[table,xcdraw]{xcolor}
\usepackage{makecell}
\usepackage{enumerate}

\usepackage{amsmath}
\usepackage{algorithmic}
\usepackage{url}

\newcommand\MYhyperrefoptions{bookmarks=true,bookmarksnumbered=true,
pdfpagemode={UseOutlines},plainpages=false,pdfpagelabels=true,
colorlinks=true,linkcolor={black},citecolor={black},urlcolor={black},
pdftitle={Anchorage: Visual Analysis of Satisfaction in Customer Service Videos via Anchor Events},
pdfsubject={Applications},
pdfauthor={Kam Kwai Wong},
pdfkeywords={video data, video visualization, customer satisfaction, visual analytics}}
\ifCLASSINFOpdf
\usepackage[\MYhyperrefoptions,pdftex]{hyperref}
\else
\usepackage[\MYhyperrefoptions,breaklinks=true,dvips]{hyperref}
\usepackage{breakurl}
\fi

\usepackage{enumitem}
\usepackage{xcolor}
\newcommand{\name}{\textit{Anchorage}}

\usepackage{xspace,xpunctuate}
\newcommand{\ie}{\textit{i.e.}\xspace}
\newcommand{\etal}{{ \textit{et al.}}}
\newcommand{\eg}{{\textit{e.g.}}}

\definecolor{visual_color}{HTML}{91B25B}
\definecolor{audio_color}{HTML}{B25960}
\definecolor{event_color}{HTML}{7D5BB2}
\definecolor{agent_color}{HTML}{BF82BB}
\definecolor{client_color}{HTML}{BBBF82}
\definecolor{pos_color}{HTML}{5ab4ac}
\definecolor{neg_color}{HTML}{d8b365}

\newcommand{\rc}[1]{\textcolor{black}{#1}}
\newcommand{\rw}[1]{\textcolor{black}{#1}}
\newcommand{\rf}[1]{\textcolor{black}{#1}}
\newcommand{\highlight}[1]{#1}

\newcommand{\visual}[1]{\textcolor{visual_color}{#1}}
\newcommand{\audio}[1]{\textcolor{audio_color}{#1}}
\newcommand{\event}[1]{\textcolor{event_color}{#1}}

\begin{document}
\author{
    Kam~Kwai~Wong,
    Xingbo Wang,
    Yong Wang,
    Jianben He,
    Rong Zhang,
    and Huamin~Qu~\IEEEmembership{}
\IEEEcompsocitemizethanks{
\IEEEcompsocthanksitem KK Wong, X. Wang, J. He, R. Zhang, and H. Qu are with the Hong Kong University of Science and Technology. Email: \{kkwongar, xingbo.wang, jhebtr, zhangab\}@connect.ust.hk and huamin@cse.ust.hk%
\IEEEcompsocthanksitem Y. Wang is with the Singapore Management University. Email: yongwang@smu.edu.sg. He is the corresponding author.}%
\thanks{Manuscript received April 19, 2005; revised August 26, 2015.}}

\markboth{Journal of \LaTeX\ Class Files,~Vol.~14, No.~8, August~2015}%
{Wong \MakeLowercase{\textit{et al.}}: Anchorage: Visual Analysis of Satisfaction in Customer Service Videos via Anchor Events}

\title{Anchorage: Visual Analysis of Satisfaction in Customer Service Videos via Anchor Events}
\IEEEtitleabstractindextext{%
\begin{abstract}
Delivering customer services through video communications has brought new opportunities to analyze customer satisfaction for quality management. 
However, due to the lack of reliable self-reported responses, service providers are troubled by the inadequate estimation of customer services and the tedious investigation into multimodal video recordings.
We introduce {\name}, a visual analytics system to evaluate customer satisfaction by summarizing multimodal behavioral features in customer service videos and revealing abnormal operations in the service process.
We leverage the semantically meaningful operations to introduce structured event understanding into videos which help service providers quickly navigate to events of their interest.
{\name} supports a comprehensive evaluation of customer satisfaction from the service and operation levels and efficient analysis of customer behavioral dynamics via multifaceted visualization views.
We extensively evaluate {\name} through a case study and a carefully-designed user study.
The results demonstrate its effectiveness and usability in assessing customer satisfaction using customer service videos.
We found that introducing event contexts in assessing customer satisfaction can enhance its performance \rf{without compromising annotation precision}.
Our approach can be adapted in situations where unlabelled and unstructured videos are collected along with sequential records.
\end{abstract}

}

\maketitle

\IEEEdisplaynontitleabstractindextext

\IEEEpeerreviewmaketitle

\IEEEraisesectionheading{\section{Introduction}\label{sec:introduction}}
\IEEEPARstart{C}{ustomer}
satisfaction, hereafter ``satisfaction," is an important service quality metric that highly correlates with the perception of brand image, loyalty, and switching behavior~\cite{background_satisfaction_oliver_2014}.
Monitoring satisfaction is advocated by several international standards (\eg, ISO 10004~\cite{intro_motivation_iso_2018}) and helps organizations of any size to evaluate their performance and obtain managerial insights. 
Satisfaction is often measured directly (\eg, self-reported surveys~\cite{intro_emotion_wong_2004}) and inferred indirectly (\eg, complaints~\cite{intro_motivation_tronvoll_2011}).
Since customers' direct responses are costly to acquire, the sample size of direct measurement is usually limited and thus damages its reliability. 
The summative post-service assessments are also weak in spontaneity and prone to cognitive biases such as leading questions and the peak-end rule~\cite{intro_motivation_peterson_1992}.
Therefore, there is a pressing need for automatic methods that evaluate satisfaction efficiently and concurrently.

Services have increasingly been provided remotely because agents can serve more clients with fewer time and location constraints.
Prior works on satisfaction analysis mainly focused on traditional digital delivery channels, such as
text messages~\cite{related_sat_app_alotaibi_2018, related_sat_app_see_2021} and phone calls~\cite{related_sat_most_park_2009, related_sat_most_ando_2020}.
These are anticipated to transform into video-based communications because of the enhanced user experience, especially for customer services~\cite{intro_service_saberi_2017}.
In the transformation, the supplement of video data has offered new opportunities for evaluating satisfaction, but it also comes with two main challenges.

First, the collected video data is in multimodalities, leading to difficulties in processing and comprehension.
Emotions have been widely adopted to model satisfaction because of their high correlation.
\rf{Therefore, previous analyses~\cite{related_sat_most_park_2009, related_sat_most_seng_2018, related_sat_most_ando_2020} have exploited multimodal emotional features to deduce the satisfaction level. 
However, emotions can have different meanings and impacts on the satisfaction dimension~\cite{intro_emotion_liljander_1997, intro_emotion_wong_2004}.
For example, a client who complains about the service demonstrates intense negative feelings and low satisfaction, but a client who smiles and says ``thank you'' might only do so out of courtesy; thus, it is difficult to infer their satisfaction.}
The problem is further exacerbated by the two-way communication setting in customer services.
Analyzing satisfaction from the client's point of view is inadequate because the agent's behaviors could be the root cause of the affective reactions~\cite{background_satisfaction_cheshin_2018}.
\rf{There is a need to investigate how to combine emotional features with other behavioral cues to analyze satisfaction more effectively.}

Second, customer services are characterized by sparse feature distributions and diverse event contexts.
While video recordings may span several minutes, the desired features (\eg, facial expressions) could last only a few seconds.
Identifying such subtle and instantaneous details from large video collections is tedious.
Furthermore, deducing the event contexts from videos requires extra attention and domain expertise.
The unstructured video data lacks a proper segmentation scheme to effectively summarize the whole service and its segments.
The sequential and temporal relationship is seldom considered for evaluating satisfaction~\cite{related_sat_most_ando_2020}, leading to an extra cost in studying satisfaction patterns.

To address these challenges, we bring forward the use of \textit{\textbf{anchors}}, \ie, semantically meaningful events that describe \rw{service procedures (\textit{operational anchors}) and observable human behaviors (\textit{behavioral anchors})}, to represent critical event characteristics.
The operational anchors introduce ordered event understanding into videos by offering sequential and temporal contexts. 
\rw{The behavioral anchors represent multimodal human behaviors compactly and straightforwardly to provide automated satisfaction evaluation.}
The two anchors were combined to navigate events of interest and interpret the progression of satisfaction levels with contexts.

We propose \textit{\textbf{{\name}}}, a visual analytics system to evaluate customer satisfaction by summarizing multimodal behavioral features in customer service videos and \rw{revealing abnormal events in service procedures}. 
{\name} generates potential operational and behavioral anchors based on a multi-perspective anomaly detection framework and provides a primitive satisfaction estimation.
A set of coordinated visualizations is designed to analyze satisfaction contextualized by anchors, such that it magnifies a conventional satisfaction score with greater sequential and temporal resolutions. 
The effectiveness of {\name} is verified through a case study and a carefully-designed user study.
\rc{We found that introducing event contexts (\ie, \textit{\textbf{anchors}}) to video analytics can enhance the performance of satisfaction evaluation tasks \rf{without compromising annotation precision}.
{\name} is useful in summarizing video contents, identifying anomalous events, and understanding multimodal features.}
Our approach can be adapted in situations where unlabelled and unstructured videos are collected along with sequential records.
In summary, our main contributions are:
\begin{itemize}[noitemsep,topsep=0pt,label=$\diamond$]

\item Problem characterization in evaluating satisfaction levels with customer service videos and machine logs through iterative discussions with domain experts.
We applied the understanding to create an improvised dataset to verify the approach's efficacy under different satisfaction scenarios.

\item A multi-level anomaly detection framework to generate anchors for efficient event understanding in video analysis and adaptation of discrete event analysis to video visual analytics.

\item Novel and metaphoric visualization designs that facilitate effective identification of multimodal anomalous events to evaluate satisfaction levels in customer service videos. 

\end{itemize}
\section{Related work}
Our work is relevant to prior studies on customer satisfaction analysis, visual analytics for multimodal videos and their event understanding, and discrete event sequence visualization.

\subsection{Data-driven customer satisfaction analysis}
Automatic satisfaction evaluation has been conducted on diverse data types, such as texts~\cite{related_sat_app_alotaibi_2018, related_sat_app_zhang_2020, related_sat_app_see_2021}, eye gaze trajectories~\cite{related_sat_app_liu_2019}, and electroencephalogram signals~\cite{related_sat_app_kumar_2019}.
These approaches rely upon a high active participation rate and specific devices that are both uncommon in many service scenarios.
Unlike other data sources, videos are more accessible, provide continuous input for in-depth satisfaction analysis~\cite{related_sat_video_gunes_2013}, and do not impose an extra cognitive burden on customers~\cite{related_sat_video_mcduff_2012}.
For example, surveillance videos in retail stores~\cite{related_sat_video_generosi_2018} and crowd-sourced web camera videos~\cite{related_sat_video_mcduff_2012, related_sat_video_mcduff_2015} have been analyzed for customer responses to products and services.

Many video-based approaches have focused on extracting facial expressions as frame-level features and evaluating overall video-level satisfaction~\cite{related_sat_video_slim_2018, related_sat_video_sugianto_2018, related_sat_video_gonzalez_2020}.
Yolcu\etal~\cite{related_sat_video_yolcu_2018} accumulated all the emotions of different customers as a proxy to estimate their satisfaction.
They also considered the head pose in tackling poor performance when the targets' faces are occluded.
\rw{However, these methods were naively evaluated on how accurately the facial expressions in the videos were identified but not the satisfaction.
They only applied a trivial model between emotion and satisfaction.}

Besides the visual channels, acoustic information is also crucial for satisfaction estimation.
Park and Gates~\cite{related_sat_most_park_2009} 
derived several prosodic and lexical features for SVM to model satisfaction.
Seng and Ang~\cite{related_sat_most_seng_2018} fused affective features in visual and acoustic channels with a linear model bespoke to satisfaction evaluation.
Ando\etal~\cite{related_sat_most_ando_2020} introduced a hierarchical framework to combine emotional features in the conversation and individual utterances.
These approaches provided data-driven satisfaction scores for videos and were tested against real and improvised datasets.
However, a video-level satisfaction score cannot identify the critical turning points and fails to distinguish the counteracted cases.
\rw{Also, these approaches put little or no emphasis on the agent's behaviors, which could be the antecedent incidents that affect the customer's consequential reactions~\cite{background_satisfaction_oliver_2014, related_sat_video_gunes_2013}.
Our work explores multimodal fusion with behavioral features and event contexts to establish background associations for interactive satisfaction analysis.}

\subsection{Visual analysis of multimodal videos}
A plethora of visualizations has been proposed to summarize and represent video data in the community~\cite{related_video_emotion_zeng_2020,related_video_emotion_ma_2020,related_video_voicecoach_wang_2020,related_video_dehumor_wang_2021,related_video_emotion_maher_2022,related_video_gesturelens_2022}.
Emotional features are commonly extracted from videos for many application problems.
EmotionCues~\cite{related_video_emotion_zeng_2020} summarized the emotional dynamics in classroom videos and highlighted model uncertainties by a stream graph design. 
EmotionMap~\cite{related_video_emotion_ma_2020} and E-ffective~\cite{related_video_emotion_maher_2022} proposed a map-based and spiral-based design to provide a temporal overview of the affective dimension in multimedia videos.
These visualizations effectively presented visual cues for notable moments.
However, videos often require more features than emotions alone to serve various domain-specific analytical purposes.

Recently, more modalities have been used to provide additional information missing from the visual channel.
EmoCo~\cite{related_video_multimodal_zeng_2021} explored the emotional coherence across facial expressions, audio emotions, and transcript sentiments to mitigate over-reliance on a particular modality. 
Li\etal~\cite{related_video_multimodal_li_2021} visualized head pose with mouse movement data to connect multimodal behaviors in online proctoring.
VideoModerator~\cite{related_video_multimodal_tang_2022} engineered risk-related features from images and transcripts to assist live stream moderation.
Besides verifying features, analyzing multimodalities has been useful in interpreting multimodal models~\cite{related_video_multiviz_2022, related_video_multimodal_wang_2022}, querying large video collections~\cite{related_video_multimodal_wu_2020}, and annotating think-aloud usability test videos~\cite{related_video_multimodal_blascheck_2016, related_video_multimodal_soure_2022}.
Although integrating more relevant data sources increases the credibility of analysis results~\cite{discussion_dataset_lin_2021, discussion_dataset_zhang_2023}, it inevitably increases cognitive loads in comprehending them.
We utilize a set of well-coordinated views to facilitate the intuitive interpretation of multimodal features.
\rf{In particular, we propose a scatter-based metaphoric visualization design to summarize the multimodal features and show satisfaction progression.}
We leverage machine logs to contextualize and grant procedural understanding to customer service videos.
It avails a new perspective in summarizing video sequences.

\subsection{Event understanding in video visual analytics}
\label{event_understanding}
According to H\"{o}ferlin\etal~\cite{related_video_survey_2015}, video visual analytics has three main goals: \textit{status determination}, \textit{event detection}, and \textit{model generation}. 
Status determination identifies frame-level features such as tracking objects.
Contrarily, the other two goals consider a larger portion of the video. 
While event detection seeks to locate the moment when a specified event occurs, model generation aims at mining common patterns from video collections that can later be used to detect events.
\rf{Our work} focuses on model generation to make sense of unlabelled satisfaction patterns. 

As the desired patterns are vaguely defined, analysts have to explore a large low-level event space to generate high-level concepts~\cite{related_video_multimodal_blascheck_2016}.
The frame-shot-scene hierarchy in movie analysis~\cite{related_video_event_kurzhals_2016} and the object-event-tactic hierarchy in sports video annotation~\cite{related_video_event_chen_2022} can be viewed conjunctively to illustrate the complexity and interdependency of video events.
\rw{
To streamline the exploration, EventAnchor~\cite{related_video_event_deng_2021} traced visually available objects in racket sports videos and denoted their critical change of states as \textit{anchors}.
These anchors were plotted on the screen to indicate the objects' locations for interactive calibration of the machine errors.
However, the computer vision-based anchors are limited in numbers and cannot be trivially applied to acoustic channels and other modalities.
}
Also, the visual effects, such as scoreboards and scene changes, are absent in many real-life video recordings to provide the event contexts.
Inspired by the anchor concept, we extend its usefulness in understanding multimodal video events and generalize it into scenarios where videos are recorded along with sequential records.
We leverage semantically meaningful events in service procedures to prioritize investigative efforts among different anchors.
We propose a semi-automatic framework that generates anchor candidates with anomaly detection methods and supports candidate validation with intuitive visualization designs.

\subsection{Discrete event sequence visualization}
Discrete event analysis usually applies to log data such as computer system records~\cite{related_event_app_shi_2014}, which have a timestamp for each event record and a semantic meaning for each event type.
\rw{
The \highlight{multi-scale temporal structure} of these events is often harnessed for visual summarization.
}
For example, event sequences can be aggregated by multivariate regular expressions~\cite{related_event_hierarchy_cappers_2018} and hierarchical clustering~\cite{related_event_hierarchy_magallanes_2022} to serve different level-of-detail requirements.
Distinguishing branches~\cite{related_event_hierarchy_liu_2017} and bundling frequent patterns in event sequences~\cite{related_event_hierarchy_polack_2018} have facilitated a further understanding of the diverging and converging event evolution patterns.
A recent survey~\cite{related_event_survey_guo_2021} has summarized the design space for event sequences.

Through multi-scale overviews, analysts can visually compare the temporal characteristics of different event sequences to locate abnormal behaviors~\cite{related_event_anomaly_nguyen_2019} and process drifts~\cite{related_event_anomaly_yeshchenko_2021}.
Guo\etal~\cite{related_event_anomaly_guo_2021} proposed a VAE-based approach to detect arbitrary ordering, absence, and duplication of events.
We borrow ideas from this line of research to discover \highlight{unusual service procedures}.
We employ a similarity-based method to find temporal anomalies and a Markovian technique to detect sequential anomalies.
They formulate operational anchors and contextualize the multimodal behavioral anchors for evaluating customer satisfaction.

\section{Problem Characterization}
This section introduces the background of satisfaction evaluation in customer services. 
We describe the requirement analysis and the details of the improvised dataset used for evaluation.

\subsection{Satisfaction and customer service}
\textit{Customer satisfaction} is widely defined as the fulfillment of customers' expectations with the perceived service quality~\cite{background_satisfaction_oliver_2014, intro_motivation_iso_2018}.
However, as suggested by the highly subjective terms ``expectation" and ``perceived," satisfaction is a complicated construct with various interpretations by different people.
Therefore, services are usually recorded to prevent misinformation and avoid conflicts in complaints, providing rich sources of recordings for analysis.

\textit{Customer service} refers to the organization's assistance and service for their customers before, during, and after-sales~\cite{intro_motivation_iso_2018}.
For example, contact centers provide customer services through phone calls for handling inquiries, managing orders, and troubleshooting issues~\cite{intro_service_saberi_2017}.
While kiosks, mobile applications, and virtual assistants have been established to let customers serve themselves, staff-assisted customer services remain irreplaceable because of regulatory requirements, business procedural complexity, and insufficient machine capabilities in directing human intents~\cite{intro_challenge_chen_2021}.

\begin{figure}[t]
	\centering
	\includegraphics[width=\linewidth]{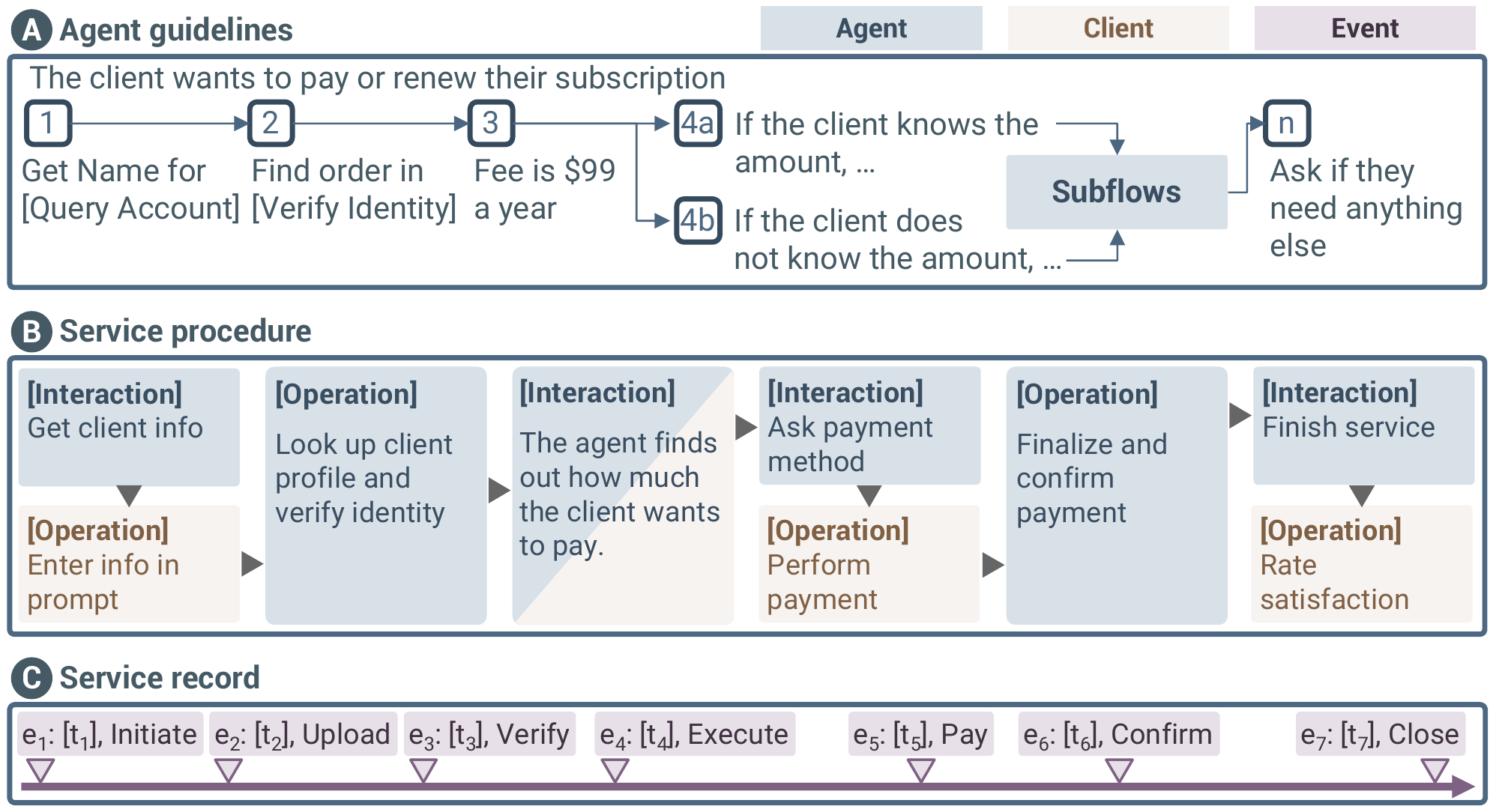}
    \caption{\rc{A customer service example modified from a customer service dialogue dataset in online shopping~\cite{intro_challenge_chen_2021}. (A) The agent guidelines outline the typical procedure. (B) \rf{The service procedure is interwoven with the interactions and operations from the agent and the client.} (C) The service record logs the executed operations with metadata such as timestamps.}}
    \label{fig:customer-service}
\end{figure}

Customer services are characterized by their goal orientation and communication dynamics.
We focus on the typical setting where a customer service agent assists a client in completing domain-specific goals.
We outline a typical workflow in~\autoref{fig:customer-service} and clarify the terms used in the paper.
First, the \textit{\textbf{agent}} assists the \textit{\textbf{client}} in specifying their intents and interpreting their needs at the beginning of a \textit{\textbf{service}}.
Second, the agent follows internal guidelines to derive a list of \textit{\textbf{operations}} and guide the client through the sequential service procedure.
Third, the agent and the client usually take \textit{\textbf{turns}} communicating and performing the operations.
\rc{For example, to verify a client's identity, the agents can ask questions verbally, prompt for entry, or make a database query for the profile.}
Finally, the executed operations formulate a \textit{\textbf{service record}} that may deviate from the agent guideline due to complex real-life situations.
The service is usually recorded to be the \textit{\textbf{service video}}.

\subsection{Design requirement analysis}
We adopted the design study methodology~\cite{background_vis_sedlmair_2012} to characterize the domain problems.
During the past two years, we have collaborated with a domestic information technology company that digitalizes public services and develops remote service terminals for tax authorities.
The terminals connect agents and clients at different locations through video communications. 
It facilitates essential operations such as transmitting legal documents and processing digital payments.
The equipment collects \textit{video recordings} and \textit{machine logs} of the staff-assisted tax filing services.

To understand their workflows and satisfaction evaluation methods, we conducted contextual inquiries and semi-structured interviews with three frontline tax officers (\textbf{E1-3}) with over 5 years of service experience.
We interviewed a professor (\textbf{E4}) with 10 years of research experience in customer relations to gain a second opinion from the marketing field.
Under their influence, we read domain literature to understand important concepts and link them with visualization studies.
We held a series of remote meetings with three business analysts (\textbf{E5-7}) with over 5 years of public service experience from our industry collaborator to refine design requirements, adapt previous findings to our application, and verify the iterative designs.
None of \textbf{E1-7} is a co-author of this paper.

Our goal was to design a satisfaction evaluation system for customer service providers to identify satisfaction patterns for improving their services and workflows. 
We identified six design requirements to support the development of {\name}. 


\begin{enumerate}[label=\textbf{R{\arabic*}}, nolistsep]
\item 
\rf{\textbf{Rank satisfaction by objective metrics.}
Clients seldom provide satisfaction feedback after services.
When they do, their self-reported evaluation is prone to cognitive biases such as the peak-end rule~\cite{intro_motivation_peterson_1992} nonetheless.
\textbf{E1-2} added, ``some clients rushed to leave, so they randomly clicked any buttons."
The large video collections also require a ranking order to prioritize videos of interest.
The system should provide uniform and objective assessments based on users’ behaviors.}

\item 
\textbf{Contextualize the satisfaction evaluation with operations.}
\rf{
Customer behaviors should be interpreted with the antecedent events~\cite{intro_motivation_tronvoll_2011}.
For example, expecting smooth services, clients would perceive repeated and interrupted operations as troublesome and unsatisfactory, resulting in a negative emotional response.
However, clients have diverse affective reactions to provocative actions.
\textbf{E7} pointed out that ``some people keep a poker face, but they could be furious," \rc{suggesting the unreliability and inadequacy of using emotional features only.}
The system should incorporate procedural considerations as the common ground to explain and evaluate clients' behaviors.}

\item 
\textbf{Show satisfaction progression in a service.}
\rc{Automatic methods usually aggregate frame-level evaluations to model satisfaction~\cite{related_sat_most_seng_2018}. 
However, the aggregated service score is inferior in differentiating counteracted cases.}
\textbf{E5} proposed a satisfied case with a client showing unsatisfied behaviors at first but becoming more satisfied with the service at last. 
The case would be underrepresented in an accumulative service-level satisfaction score.
\rc{Assessing satisfaction by individual operations naturally magnifies their contributions to the overall evaluation~\cite{related_sat_most_ando_2020}.}
\rf{The system should visualize the dynamic satisfaction progression to reveal the causal relationships between behaviors and satisfaction.}

\item 
\textbf{Provide an overview of the service record.}
\rf{The service records provide sequential and temporal information to indicate the service smoothness.
The records of smooth services usually match the typical workflows described in the agent guidelines explicitly.
Experienced agents (\textbf{E1-3}) could easily identify deviated operations when they read the records in semantically meaningful terms.
The operation duration also implicitly hints on the procedural difficulties and the service status.}
\rc{The system should present adequate event contexts to foster procedural awareness and segment the services properly.}

\item 
\rf{
\textbf{Highlight the anomalous operations.}
Satisfaction generally follows a steady progression with previous states.}
A significant turning point could indicate a potential satisfaction pattern induced by internal factors (\eg, exceeding expectations~\cite{background_satisfaction_oliver_2014}) and external factors (\eg, agents' misconduct~\cite{intro_motivation_tronvoll_2011}).
\textbf{E4} stated that looking into the ``peaks" and ``troughs" of the satisfaction level would help derive more managerial insights.
They are worth more attention to be further investigated.
The system should distinguish uncommon satisfaction development to identify critical transition moments.

\item 
\textbf{Support interactive navigation of original videos.}
Video recordings are the strongest evidence in evaluating satisfaction.
\rf{Yet, reviewing the videos from scratch is inefficient.}
\rf{Features extracted by machine learning models are helpful, but they might suffer from model uncertainty and multimodal interactions~\cite{related_video_multimodal_wang_2022}.}
\textbf{E6-7} expressed a need to validate the features when they convey ``unreasonable and contradictory meanings." 
Also, the dynamics between agents and clients are challenging to define and detect.
The system should support various interactions to streamline the fast location of interested events.
\end{enumerate}

\subsection{Improvised dataset of customer service videos}
\label{dataset}
We analyzed 20 authentic videos from \textbf{E5-7} regarding tax service assistance in local government tax authorities.
\rc{These videos had no self-reported satisfaction levels.
Most clients showed a neutral face, and the services operated on average time.
They could only reflect neutrally satisfying cases, as \textbf{E6} and \textbf{E7} provided their evaluations as ground truths.}
Obtaining more recordings was challenging because the Covid-19 pandemic lockdowns limited our presence at the local tax offices to obtain the client's consent.
To the best of our knowledge, there are no publicly available datasets that include both video recordings and service records for evaluating satisfaction.
To overcome the imbalanced distribution per satisfaction level, we created an improvised dataset for proof-of-concept.

\rc{
\textbf{Participants and apparatus.} 
We invited 26 employees (4 females and 22 males) of our industry collaborator to participate in improvising.
These participants improvised either as agents or clients.
Four business analysts were qualified as agents because they had expertise in the service workflows.
The other participants were included as clients if they had visited government authorities for staff-assisted public services.}
\rf{The recording was taken with the collaborator's terminals to simulate real illumination and occlusion settings.}
The study was approved by the internal IRB (\#HREP-2021-0162), and the videos were recorded with written consent.

\rc{
\textbf{Designs and setup.}
We designed and exemplified typical service scenarios with different satisfaction levels.
Four satisfaction types were deduced by observing the collected footage's workflows and interviewing frontline agents (\textbf{E1-3}).
They include:}
\begin{enumerate}[nolistsep]
\item[\textbf{ST}] a \textbf{S}a\textbf{T}isfied service with a shorter completion time than expected. The agent delivers clear instructions and demonstrates proficiency in completing their tasks. The client is thankful.
\item[\textbf{NM}] a \textbf{N}or\textbf{M}al service with matched expected completion time. The agent controls the time of each operation to be around average. The client is given no instructions.
\rc{
\item[\textbf{DA}] a \textbf{D}issatisfied service about the \textbf{A}gent with a longer completion time than expected. The agent demonstrates inattentive behaviors (\eg, using mobile phones and chitchatting with others) to prolong the service. The client is annoyed.
\item[\textbf{DP}] a \textbf{D}issatisfied service about the \textbf{P}rocedure with a longer completion time than expected. The service procedure is interrupted by the malfunctioned terminal, which requires the client to repeat certain operations. The client is annoyed.
}
\end{enumerate}
We expressed the satisfaction types in high-level terms as guidelines.
The participants created their own speech and reaction to improvise the services.
The agents were asked to keep a neutral face to prevent emotional contagion~\cite{background_satisfaction_cheshin_2018}.
The clients' reactions were described in abstract terms such as ``being thankful" and ``being annoyed."
\rc{We refrained from scripting the scenarios to sustain spontaneity, enhance generalizability, and avoid the curse of knowledge.}

The service scenario is about amending membership of social security insurance. 
Historical records show that the average time for the service is around 8 minutes.
Since expectation significantly impacts satisfaction~\cite{background_satisfaction_oliver_2014}, we communicated this information as the expected time to control the temporal expectation for all participants.
\rc{The typical process involves nine steps from \textit{initiate} to \textit{close}, as shown in~\autoref{fig:teaser}B1.
Our service scenario's workflow is transferable to a task-oriented dataset about customer service~\cite{intro_challenge_chen_2021}.}

\rc{\textbf{Result.}}
\rc{We collected 61 service videos, with at least twelve for each satisfaction type.
We ensured no clients repeated acting in the same type.
The total duration of the services is 5.8 hours, and each spans 3-12 minutes, averaging 5.7 minutes.
The corresponding satisfaction type labels the ground truth of the videos.}

\section{Anchor generation}
\rf{
This section introduces the construction of anchors.
Anchors refer to \textit{semantically meaningful events} that describe operations (\textit{operational anchors}) and behaviors (\textit{behavioral anchors}).
We identified anomalous events for services and operations as operational anchors with a multi-perspective anomaly detection framework (\textbf{R5}).
We further extracted multimodal features from service videos to compute primitive satisfaction estimations for services and operations as behavioral anchors (\textbf{R1}).
These anchors can be viewed as an interactive table of content to define the video event structure (\textbf{R2}) for quick navigation to desired segments of satisfaction patterns without searching the whole video (\textbf{R6}).
}
\subsection{Processing multimodal features}
\label{multimodal_features}
We processed the visual and audio channels decomposed from videos separately and aligned them with the parsed machine logs.

\textbf{Visual features.}
We detected the bounding boxes of faces on every frame with YOLO5Face~\cite{method_visual_qi_2021} and applied triangular smooth to reduce glitches.
Since occlusion is still challenging for facial expression recognition (FER)~\cite{method_visual_li_2020}, we detected the head pose with FSA-Net~\cite{method_visual_yang_2019} to validate the reliability and reduce the impact of misclassification.
\rf{We adopted a FER model~\cite{method_visual_wang_2020} and aggregated the output discrete emotions into three large classes (positive, neutral, and negative) because the correspondence between discrete emotions and satisfaction is unclear~\cite{background_satisfaction_oliver_2014, related_sat_video_yolcu_2018}.}
\textbf{E5-7} were confused with the role of sadness and fear in evaluating satisfaction in customer services.
Moreover, sacrificing granularity for generalizability is a common approach in the affective analysis~\cite{related_sat_video_gunes_2013}.

\textbf{Audio features.}
We applied a speaker diarization algorithm~\cite{method_audio_bredin_2020} to remove noise, locate speech segments, and cluster utterances by speakers.
\rc{Since there are two speakers in a video, we registered the agent with heuristics, such as identifying the common speaker across two videos with the same agent.
The audio segments are piped into an audio emotion classification model~\cite{method_audio_pinto_2021}.}
The discrete emotion outputs are also aggregated, as in facial expressions.

\begin{figure*}[t]
    \centering
    \includegraphics[width=0.98\linewidth]{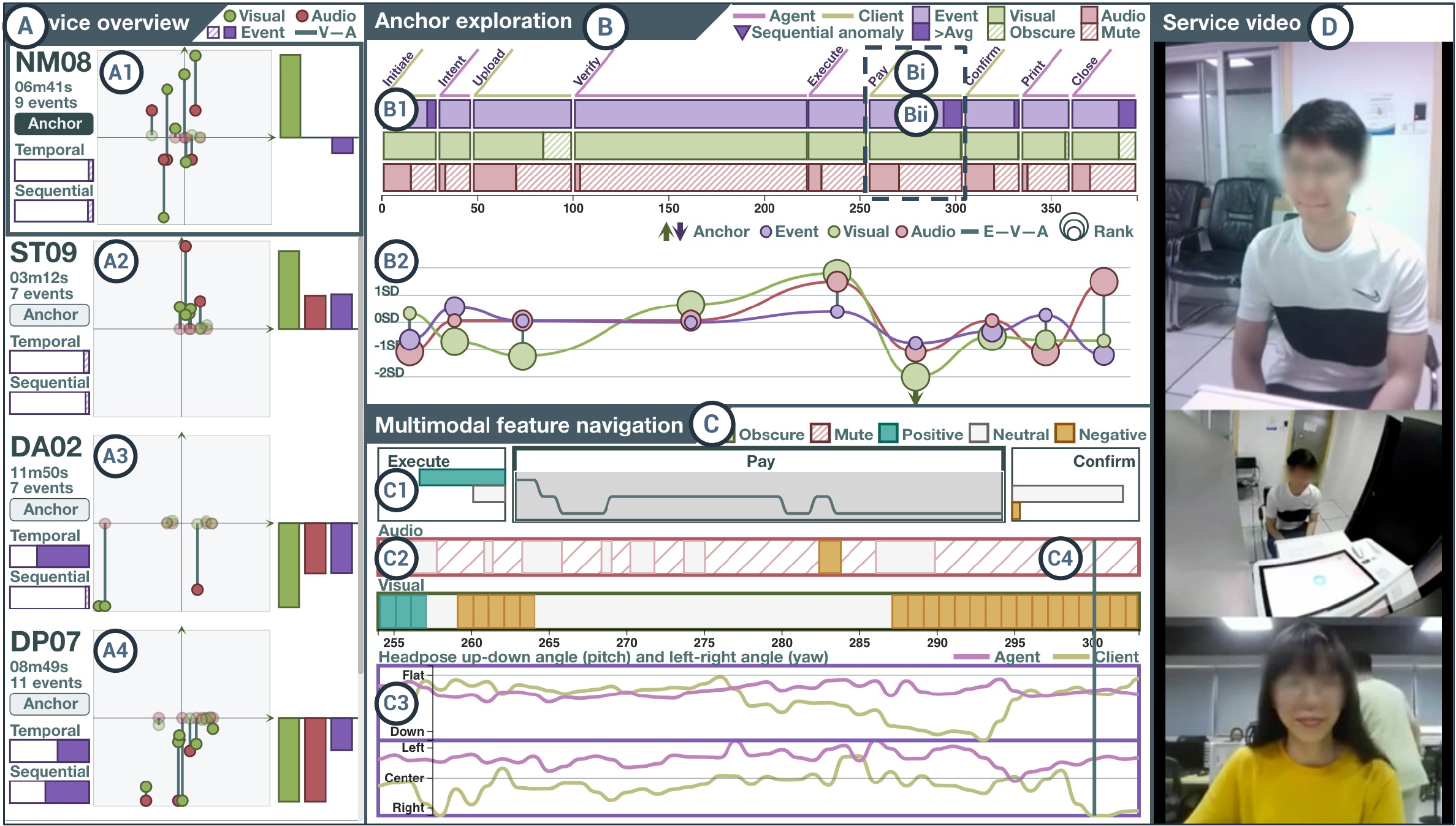}
    \caption{The interface of {\name} magnifies customer satisfaction metrics with greater sequential and temporal resolutions. The service overview (A) shows the primitive satisfaction scores based on clients' responses and the smoothness of service procedures. The anchor exploration view (B) displays the customer satisfaction progression in a service segmented by operations and highlights anomalies. The multimodal feature navigation view (C) provides detailed multimodal information for verifying the satisfaction scores and supports interactive navigation to the service video view (D).}
    \label{fig:teaser}
\end{figure*}

\textbf{Event features.}
\rc{
The machine logs contain discrete events that describe operations in the service records.
However, an operation could be represented as multiple unstructured free-form text messages due to inconsistent coding styles.}
We first transformed the logs into tuple representations that contain a timestamp, an event type, and a list of log parameters for analysis.
A service is identified by matching a beginning and an ending log message with the same terminal request ID.
\rc{We aggregated the co-occurring raw event types that are semantically related, and confirmed the aggregated event types, denoted as operations, with \textbf{E5}.
The nine operations are used to segment the videos (\autoref{fig:teaser}B1).
We counted the logs with the same operation $e$ to obtain a service record vector $E = [(e_1, n_1), (e_2, n_2),...]$, where $n_i$ is the count of consecutive $e_i$.}

\subsection{Operational anchors}
\label{Sec:operation_anchors}
\begin{figure}[t]
	\centering
	\includegraphics[width=\linewidth]{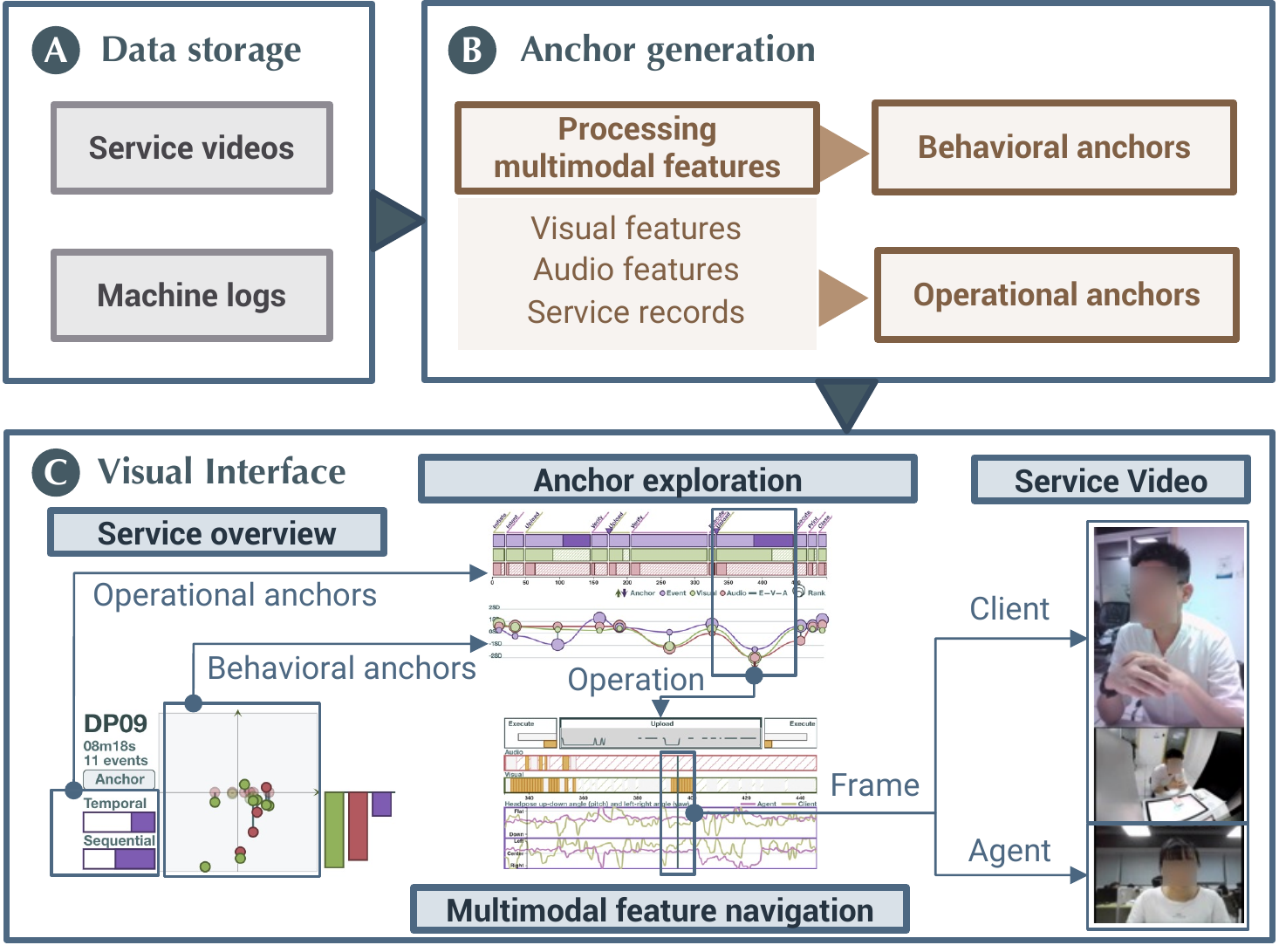}
    \caption{The system architecture of {\name} contains three modules: the data storage, the anchor generation module, and the visual interface.}
    \label{fig:framework}
\end{figure}
The primary purpose of operational anchors is to lift the burden of status determination (as discussed in \autoref{event_understanding}) for analysts when they watch the videos.
\rc{The operational anchors can segment operations in service records and grant semantic meanings to segments (\textbf{R4}).
They also introduce event structure to summarize video content.
To prioritize anomalous events (\textbf{R5}), we employed a similarity-based method to find temporal anomalies and a Markovian technique to detect sequential anomalies from the service records.
The service record vector $E$ is piped into the following algorithms to obtain the corresponding anomaly scores.}

\textbf{Temporal anomaly} locates uncommon durations of operations.
The Principle Component Analysis (PCA)~\cite{method_anchor_xu_2009} is a popular unsupervised method for system log analysis detecting anomalous discrete events.
\rf{It computes the similarity between \rc{the input and the labeled sequences} based on the assumption that anomalous sequences should be dissimilar to normal ones.}
\rf{Service records labeled as normal $E_n$ are further aggregated by the operations to obtain fixed-size vectors.
They are reduced to $k$ principal components to formulate the normal space $S_n$.}
A service record is said to be anomalous if $||y||^2 > Q_{1-\alpha}$, where $y$ is the projection length to $S_n$, and $Q_{1-\alpha}$ is the confidence threshold defaulted at 95\%.
We had considered another popular method in system log analysis, invariant mining~\cite{method_anchor_lou_2010}.
However, it is tailored to rigorous procedures in software systems and has limited generalizability.
Meanwhile, PCA has the advantage of high interpretability and does not require a large training set.

\textbf{Sequential anomaly} locates uncommon chronological orders in service records.
The Markov chain model~\cite{method_anchor_ye_2000} learns a transition probability distribution $P$ of different discrete states at each time frame in normal sequences.
It assigns the service vector $E$ with:
\begin{equation}
\label{markov}
    P(E) = P(e_1, e_2, ..., e_T) = \prod^{T-1}_{t=1}{p_{e_{t},e_{t+1}}}
\end{equation}
where $T$ is the fixed window size.
\rf{A non-zero constant $\epsilon$ is introduced to prevent zero probability when a particular sub-sequence has not appeared in the training set.
We scaled up and sampled down the event records to create fixed-sized event vectors.}
We set the anomalous threshold for operations at $|1/n|$ such that the transition occurs at least once in the training set and for service at a constant that captures 95\% variance.
The Markov chain model is chosen for three reasons:
(1) It does not require large training data;
(2) It has good scalability by enlarging the window size to facilitate a large number of log records;
(3) It identifies the exact operation when it deviates from standard procedures (\textbf{R4}).
\rf{Moreover, customer services have predefined procedures (agent guidelines) acting as the normal training set.
Although the Markovian model is not designed to identify missing and abundance events~\cite{related_event_anomaly_guo_2021}, its anomaly score would still reflect these conditions as they would appear in the wrong place. 
The Markovian model is well-suited to detecting repeated and out-of-sync operations (\textbf{R2}).}

\label{buoy}

\subsection{Behavioral anchors}
\label{Sec:behavioral_anchors}
The behavioral anchors are the multimodal satisfaction evaluation.
Similar to~\cite{related_sat_video_yolcu_2018, related_sat_most_seng_2018}, we adopt a linear model to generate a customer satisfaction score.
We extended the model to cover event duration rather than affective status only (\textbf{R2}).
The model combines facial expression $v$, audio emotion $a$, and events $e$ to evaluate satisfaction. 
The customer satisfaction score for a service $CS_{s}$ is calculated by:
\begin{equation}
\label{cs_score}
    CS_{s} = w_{v}f(\sum_{i=1}^{N}m_{v}v_{i}) + w_{a}f(\sum_{j=1}^{M}m_{a}a_{j}) - w_{e}\sum_{t=1}^{T}z_{e, t}
\end{equation}
where $N$, $M$, and $T$ denote the total number of frames, utterances, and operations.
\rc{$w$ is the weights of each channel defaulted as equally weighted.
We also obtained the operation's satisfaction score $CS_{e}$ for each modality by confining the summation scope to individual operation and modality.}
$f$ is the normal standardization across all services.
$z_e$ is the z-score for the event duration.
We grouped them by operations before standardizing because repeated operations are usually shorter and obfuscate the calculation.
For emotional responses, we assigned \rc{a magnitude weight $m$} to each discrete emotion and adopted the scheme proposed by previous work~\cite{related_sat_most_seng_2018}.
In general, positive emotion has a value of +1.0, neutral emotions are 0.0, and negative emotions tend to -1.0.
We slightly modified the weightings of anger to -1.2 and disgust to -1.0 based on the domain literature~\cite{intro_emotion_liljander_1997} and discussions with \textbf{E5-7}.
A large value of $CS_s$ indicates high satisfaction and vice versa.
All of the above settings can be reconfigured to adapt to other needs. 



\section{Visual interface}
The visual interface of {\name} supports satisfaction evaluation at multiple scales and anchor candidate validation with intuitive visualization designs.
\autoref{fig:teaser} shows the snapshot of the interactive system annotated with (A) the service overview, (B) the anchor exploration view, (C) the multimodal feature navigation view, and (D) the service video view.
\rc{The service video view shows the original service videos and plays them in sync with other views when the corresponding video or event is selected.}
It also supports conventional video playback functions and other interactions described in the following sections.

\begin{figure}[t]
	\centering
	\includegraphics[width=\linewidth]{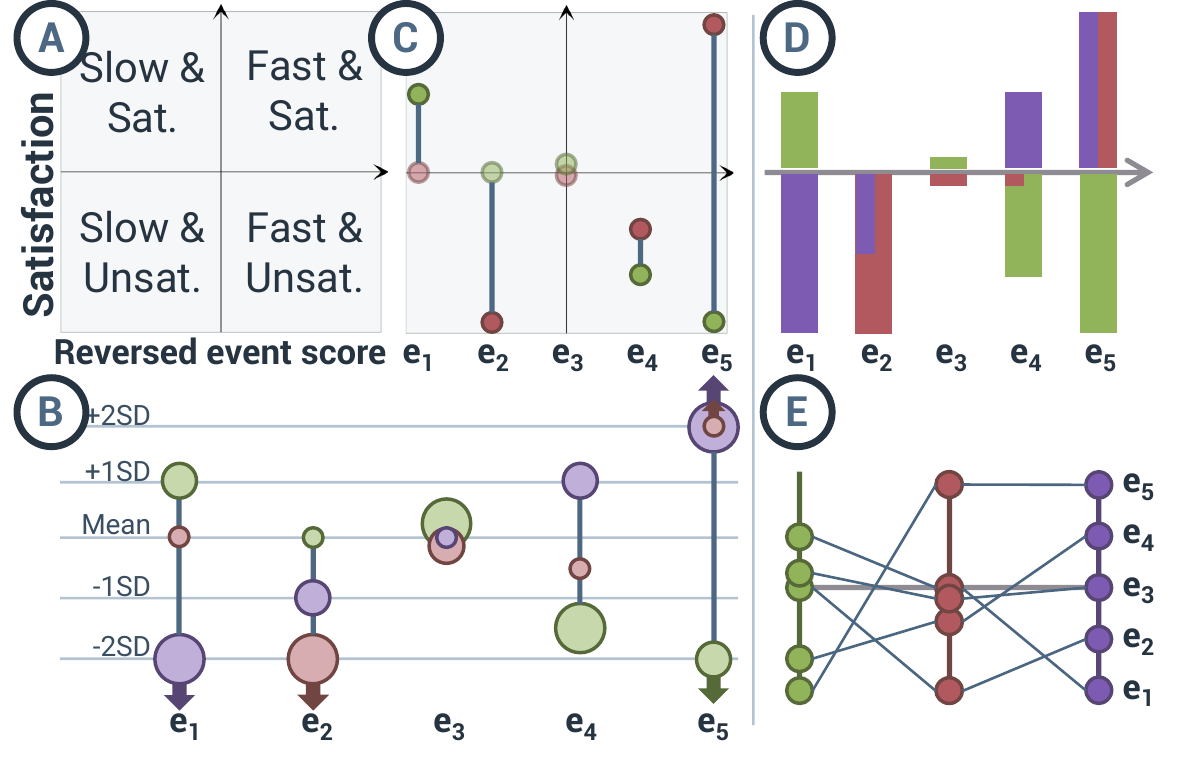}
    \caption{\rc{{\name} employs the buoy metaphor to summarize multimodal satisfaction evaluation with procedural understanding. (A) shows the four quadrants heuristic of the buoy chart. 
    (B) is the lateral buoy chart that progressively plots the multimodal satisfaction scores for the five operations ($e_{1-5}$). (C) is the buoy chart that summarizes all operations of a service in a single graph. (D-E) are the alternative designs of (C).}}
    \label{fig:buoy}
\end{figure}

\subsection{Service overview and the buoy chart}

\rc{The Service overview (\autoref{fig:teaser}A) provides an overview of all the service videos.
It lists all the videos and supports fast comparison over multiple videos to search for a service of interest.
Each list item (\autoref{fig:teaser}A1) contains three columns that show different satisfaction metrics (\textbf{R1}).
The color encodings are unified for the visual interface (\visual{green for visual}, \audio{red for audio}, and \event{purple for event}).
The leftmost column displays the basic information of the video.
The horizontal bar chart shows the temporal and sequential anomaly scores described in \autoref{Sec:operation_anchors}.
Identified anomalies are represented by \event{filled color}, and normal services are in \event{striped color}.
The rightmost column is a vertical bar chart showing the satisfaction scores $CS_{s}$ of different modalities described in \autoref{Sec:behavioral_anchors}.}

The middle column is a scatter-based design called the \textit{buoy chart} (\autoref{fig:buoy}C).
It summarizes the multimodal satisfaction scores of individual operations into a single graph.
An operation is represented by two dots, which are in different colors to indicate their types (\visual{visual} or \audio{audio}).
The vertical position encodes the dot's satisfaction score $CS_{e}$ of its type.
The dots close to the horizontal line have lower opacity, making them visually insignificant.
\rf{The horizontal position encodes the reversed event score for all dots.
The negation is designed to match the quadrants heuristic by converting a negative score, which denotes a shorter duration, into a positive.}
Moreover, two dots belonging to the same operation will be linked vertically to show a connected visual component based on the principle of continuity.
The vertical and horizontal scales are centered at zero and capped within a threshold.

\rc{
\textbf{Justification.}
We developed a metaphor to flatten the buoy charts' learning curve and lower the bar of visualization literacy~\cite{visual_design_yang_2021}.
The buoy chart employs the buoy metaphor as a buoy attaches to an anchor.
A dot is referred to as a buoy.
Normal services should have many ordinary operations; thus, dots are clustered around zero.
While these buoys in proximity float on the surface imperceptibly, the significant buoys sink and rise to become visually apparent anchors.
These outliers highlight anomalous operations with potential satisfaction patterns (\textbf{R5}).
They detect counteracted cases (\textbf{R3}) when the polarized anchors are seen together with the near-zero aggregated scores (\eg, the audio channel of \autoref{fig:teaser}A1).}

\rc{
Moreover, the buoy chart can be interpreted with the quadrant heuristic in \autoref{fig:buoy}A.
We estimate the overall satisfaction by checking the quadrant with the most anchors.
For example, \autoref{fig:teaser}A3 shows many floating buoys and a few sinking anchors.
The anchors' positions suggest that the client exhibited negative emotions in a few prolonged operations.
It shows how the buoy chart efficiently summarizes the distinguishing patterns (\textbf{R1}).
The buoy chart is the visual alternative for multimodal fusion (\eg, \autoref{cs_score}) which may require extra effort to optimize the cost functions.}

\rc{
The buoy chart effectively encodes multimodal characteristics. 
\autoref{fig:buoy}C demonstrates the visual patterns of the three inter-modal interaction types summarized by Wang\etal\cite{related_video_multimodal_wang_2022} ($e_{1-3}$ for dominant, $e_4$ for complementary, and $e_5$ for conflicting modals).
We use the four cases in \autoref{fig:teaser} as examples:
A2-A3 dominantly express strong emotions in one channel; A4 conveys negative emotions by complementing both channels; A1 shows conflicting behaviors for some operations.
These visual cues help analyze the clients' emotional profiles (\textbf{R2}), which inform the subsequent analysis.} 

The buoy chart can be augmented to address various design issues. 
Techniques applicable to scatter-based designs are also likely effective for the buoy chart.
For scalability issues, we can reduce the dot size and superpose bar charts to the sides to observe the operation's distributions in densely populated regions.
To prioritize the most important items, we can use a quadrant-based heatmap to filter and rank the videos by anchor patterns of interest.

\rc{\textbf{Design alternatives.} 
We focused on designing straightforward and standardized charts to suit the diverse background of our target users, \ie, service providers and agents.
Building abstractions is a popular strategy for handling numerous videos in video collections.
We implemented dimension reduction techniques to generate video clusters and visualize outliers to reduce review efforts.
However, the techniques neglect temporal relationships and could not detect counteracted cases.
Although we can set up exemplars to guide the clusters, it is challenging to make novices aware of the technical assumptions and avoid over-reliance on unsupervised results.
We used the ``view sequentially" strategy~\cite{visual_design_gleicher_2018} instead and provided the satisfaction scores in different modalities as the ranking orders.}

As an alternative to the buoy chart, we considered using the stacked bar chart to display the operations sequentially, as in \autoref{fig:buoy}D.
It is visually apparent when the multimodal scores are dominant.
However, the complementary and conflicting modalities challenge the interpretability of the chart for lacking quick decision rules.
The chart also creates confusion when displaying positive and negative values together, and suffers scalability issues with more operations.
Another option was the parallel coordinates (\autoref{fig:buoy}E).
However, the stronger intra-modal scalability cannot compensate for the visual clutter of lines when performing inter-modal comparisons. 
It is also complicated to compare three modalities simultaneously to find anomalous operations. 
\autoref{fig:buoy}C-E share the same set of data to highlight their difference.

\subsection{Anchor exploration view}
\rc{This view (\autoref{fig:teaser}B) supports operation-level anchor exploration based on its service record.
The timeline-based visualization (\autoref{fig:teaser}B1) represents each operation with a column of visual components.
The horizontal position encodes the service time.
Each column contains four rows.
In the top row \rf{(\autoref{fig:teaser}Bi)}, we visualize the operations, turn-taking information, and the indicator of sequential anomalies to provide the procedural context and indicate sequential inconsistency (\textbf{R4}).
The line is colored pink for the agent's turn or yellow for the client's.
The triangle icon indicates that the operation is sequentially anomalous, as in \autoref{fig:case2}A.}

\rc{
The bottom three bars \rf{(\autoref{fig:teaser}Bii)} illustrate the statistics around the event, visual, and audio modalities\rf{, respectively}.
The \rf{first bar in purple} shows the duration of the operation. 
The portion in dark purple indicates the amount of time exceeding the operation's average.
\rf{It signals a longer-than-usual operation and can be considered a temporal anomaly.
The second bar in green and the third in red represent the proportion of time with detected features for the visual and audio channels.}
The striped pattern encodes the absence of features, such that green is for obscured client's face and red is for silence.
\rf{For example, \autoref{fig:teaser}Bii shows that the client's face is not obscured for the whole operation, and the conversation lasts for about one-third of the time. }
These indicators provide background information about the reliability of the detected features.
The visual components are associated with other views, so clicking on them can navigate to the multimodal features and the original service videos (\textbf{R6}).}

\rc{The \textit{lateral buoy chart} (\autoref{fig:teaser}B2) shows the satisfaction progression.
The chart's horizontal encoding follows the timeline above, so all dots are located in the middle of the operation.
Its correspondence with the buoy chart is illustrated in \autoref{fig:buoy}B-C. 
While the two charts share the same metaphor, there are subtle differences.
Each operation is represented by three dots, including the event.
Here, the vertical position utilizes the z-score to unify all modalities.
The visual and audio scores are summed over the operation and further standardized within the selected service.
For example, $e_5$ in \autoref{fig:buoy}B contains vastly deviated scores for all modalities, while the $e_3$ counterparts have average scores.
An anchor icon denotes higher values than two standard deviations. 
The scale helps detect the most anomalous service operations (\textbf{R5}).
The buoy's size encodes the absolute deviation rank to draw attention to the most influential anchor. 
The more significant deviation, the larger the buoy.
The drawing order favors smaller buoys to prevent occlusion and visual clutters (see \autoref{fig:case2}A).}

\rc{
\textbf{Justification.}
The lateral buoy chart bridges the gap between the buoy chart and the timeline. 
We did not explicitly encode the operations shorter than average in the timeline because they could distort the layout.
Also, they are less significant as shorter events usually have fewer behavioral anchors to verify.
The missing temporal anomalous information is covered in the lateral buoy chart with the introduction of event buoys.
Using familiar visual elements and metaphor reduce the burden of learning a new visualization.
The correspondence could introduce higher efficiency when users are familiar with the system.}

\rc{
\textbf{Design alternatives.} 
During the design process, we referred to the event sequence design space~\cite{related_event_survey_guo_2021} and quickly eliminated hierarchy-based, Sankey-based, and matrix-based designs because of the unfit tasks.
We created the current design by combining bar charts and timelines for simplicity and familiarity.
For visualizing feature progression, Zeng\etal\cite{related_video_emotion_zeng_2020} proposed five designs to show emotion flows.
However, these designs are limited in visual summarization power because they lack event contexts.
Our lateral buoy chart combines multiple visual elements to coherently express the satisfaction progression in service operations.}


\subsection{Multimodal feature navigation view}
\rc{This view supports interactive navigation of the original videos (\textbf{R6}).
We adopted the periphery plot~\cite{visual_multimodal_morrow_2019} as the operation summary (\autoref{fig:teaser}C1).
In the middle focused detail view, we fused the facial and audio features to assign an activation value $v_i = \{-1, 0, 1\}$ to frame $i$.
The fusion favors non-neutral emotions with higher priority given to negative ones because they have a greater impact on satisfaction~\cite{intro_emotion_liljander_1997}.
The activation values are plotted to show an overview of the operation.
Brushing selects the period for the below features.
The periphery plots on both sides allow quick navigation to consecutive operations and contextualize the focused operation with neighbors.
The three bars show the count of activation values.}

\rw{The audio and visual channels are one-dimensional shaded matrices (\autoref{fig:teaser}C2) that encode the positive and negative outputs of facial expressions and audio emotions.
Obscured and muted frames are encoded with the stripped pattern as before.
The head pose information is shown in line charts (\autoref{fig:teaser}C3).
The chart scales are written with the semantic meaning of the angles directly.
The proximity between visual features and head pose allows users to verify emotions with occlusion from looking down.
Clicking on the views (\autoref{fig:teaser}C4) seeks the time point in the video.}

\section{Evaluation}
\rc{We present a case study with \textbf{E5} and a structured user study to demonstrate the effectiveness and usability of {\name}.}
\subsection{Case study}
\begin{figure}[t]
	\centering
	\includegraphics[width=\linewidth]{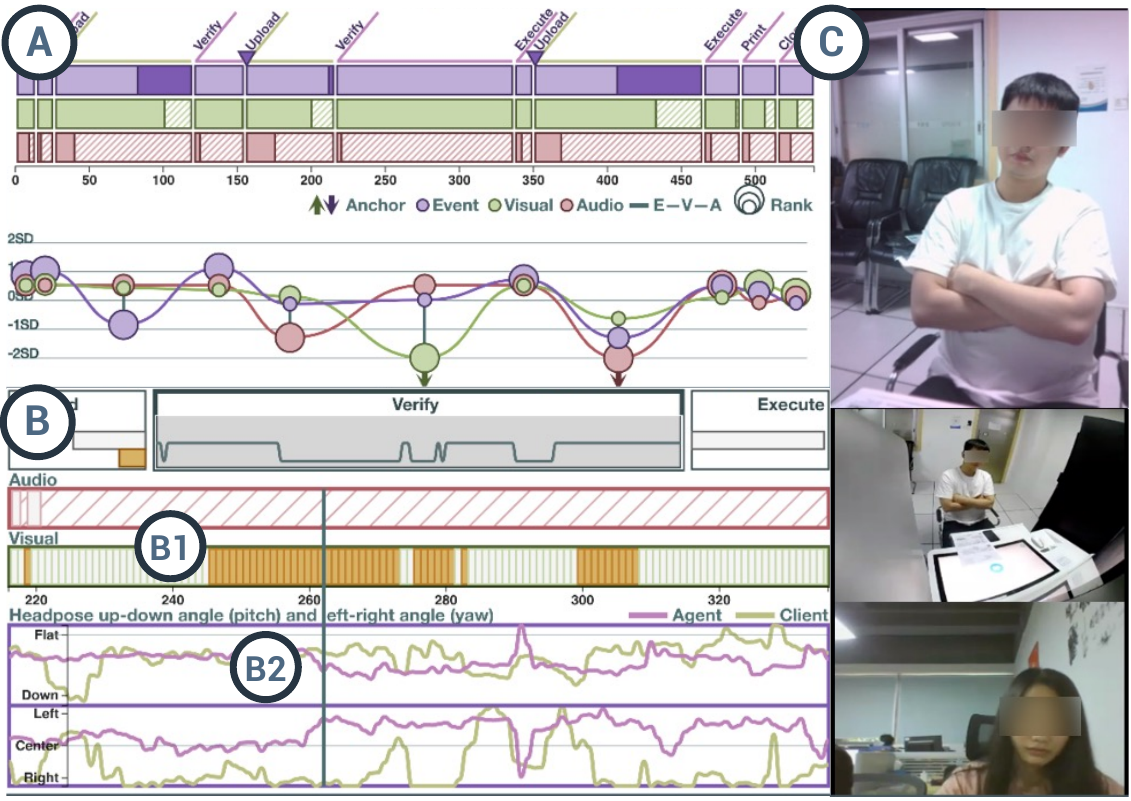}
    \caption{The unsatisfied case shows a client feeling annoyed by the repeated operations and the inattentive agent.}
    \label{fig:case2}
\end{figure}

\rc{This study describes the satisfaction evaluation process of \textbf{E5} for services provided by two of his actual subordinates, \textbf{S1} and \textbf{S2}. 
He was tasked with rating the services and the agents by exploring the improvised dataset (\autoref{dataset}) filtered by \textbf{S1} and \textbf{S2}.
He has been closely involved in the problem characterization and improvised dataset formulation, but he had not used {\name} nor seen the videos before the case study.
He knew about the satisfaction scenarios typically known to frontline agents.
He was encouraged to follow the think-aloud protocol during his exploration.}

\rc{\textbf{Service exploration (R1-2).} 
Beginning from the service overview (\autoref{fig:teaser}A), \textbf{E5} first ranked the videos by descending service satisfaction score.
ST09 (\autoref{fig:teaser}A2; Video names were masked during the exploration) was ranked number one on the list. 
He noticed that all metrics favored the service because all bars showed positive values, and there was a distinctive audio anchor in the buoy chart.
Reading the basic info on the left column, he quickly determined it should be the \textbf{ST} type.
He selected the service to see who the agent was (\textbf{S1}) from the service video view.
Then, he continued browsing other videos.
After a few attempts, his attention was caught by NM08 (\autoref{fig:teaser}A1).
He examined the near-zero audio satisfaction score and the polarized buoys in the buoy chart and suspected it was a counteracted case.
He felt interested in the case and wanted to know about the conflicting behaviors.
He selected the service and proceeded to explore the service record.}

\rc{\textbf{Operation exploration (R3-5).}
Looking at the anchor exploration view (\autoref{fig:teaser}B), he first checked the procedure summary (\autoref{fig:teaser}B1) and did not find many sequential or temporal anomalies.
Most of the agent's operations finished in time, and no procedures deviated from agent guidelines.
He ruled out the \textbf{DP} type.
He turned to the lateral buoy chart (\autoref{fig:teaser}B2) and discovered the visual anchor indicating significant negative facial expressions.
He also noticed that the previous operation of the anchor was very positive. 
Revisiting the operations' names (``Execute" and ``Pay"), he had a clue about the incident but needed more evidence.
He clicked on the anchor icon to investigate the critical transition moment.}

\rc{\textbf{Feature verification (R5-6).} 
Entering the multimodal feature navigation view (\autoref{fig:teaser}C), he found that the most negativity was located in the latter part of the ``Pay" operation from the periphery summary and the visual feature (\autoref{fig:teaser}C4).
He clicked on the orange frames at C4 to navigate the service videos (\autoref{fig:teaser}D).
By watching the original video, he concluded that the negative emotions came from having to pay but not because the agent was inattentive.
He rejected the \textbf{DA} type and declared it the \textbf{NM} type.
However, he wondered why the client had many positive behaviors during the previous operation as he read the left periphery plot (\autoref{fig:teaser}C1).
He clicked on the plot and repeated the feature verification analysis.
He was intrigued by the fact that the client was only texting on his phone the whole time. 
This reinforced his \textbf{NM} rating, despite the high satisfaction score on the visual channel.
He became confident about {\name}'s ability to detect counteracted cases.
\textbf{S1} accrued a few \textbf{NM} and \textbf{ST} cases to be rated as good performance.
}

\rc{\textbf{Satisfaction evaluation (R1-6).} 
By ranking the videos in ascending scores, \textbf{E5} found DP07 (\autoref{fig:teaser}A4) to have the lowest satisfaction score in both visual and audio channels.
He noticed that the service was detected with both temporal and sequential anomalies.
He reviewed the service in the anchor exploration view (\autoref{fig:case2}A). 
He noticed that the client took longer than usual to upload his files. 
The procedure was also repeated twice such that it could annoy the client.
He concluded that the service belonged to the \textbf{DP} type because he knew it was the only scenario.}

\rc{\textbf{A DAP case.} 
The visual anchor under the ``Verify" operation drew \textbf{E5}'s attention.
He was confused because the typical \textbf{DP} scenario does not stage like this.
He investigated the visual anchor and saw the lasting orange frames in \autoref{fig:case2}B1.
From the head pose chart, he observed that \textbf{S2} had been looking down.
He clicked on the behavior (\autoref{fig:case2}B2) and derived from the videos that \textbf{S2} had been playing on her phone (\autoref{fig:case2}C).
He speculated \textbf{S2} might have combined the \textbf{DA} and \textbf{DP} types to create a more unsatisfied \textbf{DAP} case which annoyed the client with the prolonged procedure and the inattentive agent. 
Nevertheless, he rated \textbf{S2} as having poor performance as an agent, but good performance as a business analyst in taking the initiative to create new requirements.}

\subsection{User study}
\begin{figure}[t]
	\centering
	\includegraphics[width=\linewidth]{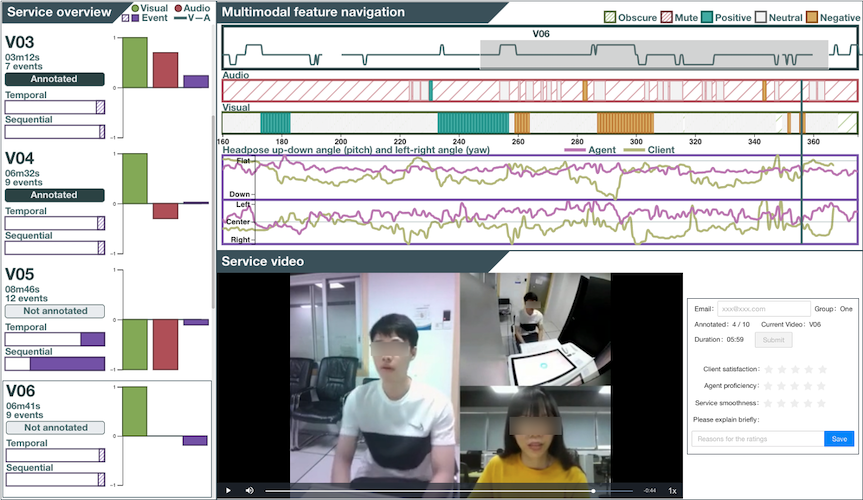}
    \caption{\rc{The interface of the baseline system used in the user study. It visualizes the customer satisfaction scores and multimodal features for basic computational supports. It is modified from {\name} by removing the event-based visual components to isolate their effects.}}
    \label{fig:baseline}
\end{figure}
\rc{We conducted the user study with a between-subject design and two conditions on the system used to evaluate the effectiveness and usability of {\name} in evaluating customer service videos.}

\rc{
\textbf{Apparatus.}
Presenting multimodal features in a VA system generally leads to better task performance than a baseline system without much computational support~\cite{related_video_multimodal_li_2021, related_video_event_deng_2021, related_video_multimodal_tang_2022}.
However, it could be unfair to compare {\name} with both operational and behavioral anchors to a naive baseline because of the wide interaction gap and the compound effect.
In this study, we aimed to evaluate whether the introduction of event context (\ie, operational anchors) enhances the performance of satisfaction evaluation tasks.
\rf{We created the baseline system (\autoref{fig:baseline}) by ablating the event-based visualization components in {\name}, namely, the buoy chart, the anchor exploration view, and the periphery plots.}
The remaining parts visualize multimodal features beneficial to automated satisfaction evaluation~\cite{related_sat_most_seng_2018, related_sat_most_ando_2020}.
The user's mouse actions were logged for provenance analysis.
}

\begin{figure}[t]
	\centering
	\includegraphics[width=\linewidth]{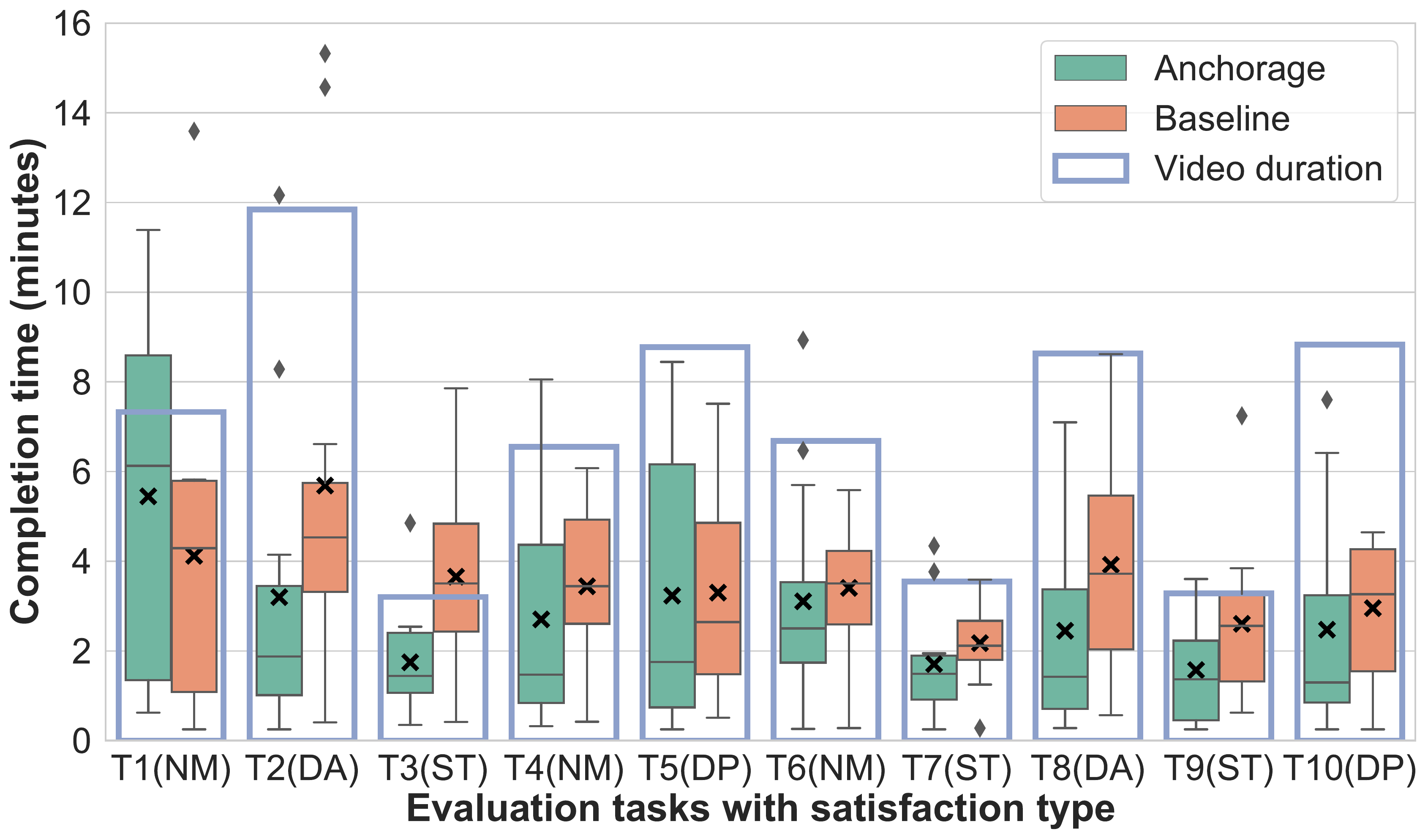}
    \caption{The completion time for each task and the duration of the corresponding videos. $\times$ represents the average completion time.}
    \label{fig:time}
\end{figure}

\textbf{Data and tasks.}
We sampled videos from the dataset described in \autoref{dataset} for the satisfaction evaluation tasks.
From a pilot study with \textbf{E5-6}, we estimated that participants could annotate three videos in ten minutes.
We randomly selected 12 videos for the study in consideration of workload and duration. 
The videos are selected with two constraints: 1) equal coverage of all four satisfaction types, and 2) acted by different clients.
They ensure the samples' diversity and independence.
We used two videos for demonstration, and participants evaluated the remaining.
The remaining ten videos span 68.6 minutes and formulate the ten satisfaction evaluation tasks (\textbf{T1-T10} in \autoref{fig:time} \rf{with masked video names}).
In addition to rating the client satisfaction on a Likert scale from 1 (low) to 5 (high), we asked them to evaluate the agent proficiency and the service smoothness to further distinguish the videos.

\rc{
\textbf{Participants.} 
We adopted snowball sampling starting from the colleagues of \textbf{E5-7} to recruit 24 participants from our collaborators' company (8 female, 16 male; age: Mean $(M) = 28.4$, Standard Deviation $(SD) = 5.3$).
While 16 participants are undergraduates in STEM disciplines, others attain diplomas in diverse backgrounds.
They have 1-13 years of related experience in customer services ($M = 5.1$, $SD = 3.7$).
They were compensated with \rf{\$7.50 USD} equivalent upon completion.
We randomly assigned the participants \rf{to use the \textit{Anchorage} system (\textbf{$P_A$}) and the \textit{Baseline} system (\textbf{$P_B$}).}}

\textbf{Procedure.} 
\rc{
The study was conducted remotely due to quarantine restrictions.
The participants could access the assigned system deployed online.
We first obtained their consent, and introduced the research background and the system's functions via recorded videos for around 12 minutes.
After three minutes of free exploration with the training examples, the participants should complete the ten tasks using the assigned system.
Since we did not enforce a time limit on the tasks, we provided cash incentives to prevent low-quality responses.
Each satisfaction rating is \$1.5 USD (maximum five rewards), if it matches the ones by \textbf{E5-6} within one point scale.
Finally, they filled in a questionnaire about the assigned system and their background information.
All sessions spanned between 30-90 minutes ($M = 56.5$, $SD = 18.3$).
}

\subsubsection{Results}
\rc{
The \textit{Anchorage} group had a shorter completion time than the Baseline group, while the annotations mainly stayed the same. 
}

\rc{
\textbf{Completion time.}
We compared the completion time in minutes of the Anchorage group ($A$) and the Baseline group ($B$) on the ten evaluation tasks.
Using Anchorage ($M_A = 27.6$, $SD_A = 19.3$) is, on average, 21.8\% faster than using Baseline ($M_B = 35.3$, $SD_B = 17.1$) in evaluating the service videos, although the Mann-Whitney U test suggested that the difference is not statistically significant (W = 51.0, p = 0.118).
Two individual tasks were significantly faster for the Anchorage group, namely, T2 ($U = 38$, $p < 0.05$) and T3 ($U = 30$, $p < 0.01$).
Combined with the findings in the case study, the two satisfaction types (\textbf{DA} and \textbf{ST}) could be easier for Anchorage users to make preliminary decisions under event contexts.
From \autoref{fig:time}, we observed that the Anchorage group has a shorter average and median completion time for all tasks except T1.
Since T1 was the first task, a possible reason is that the time needed to learn the Anchorage system is longer than that of the Baseline.}

\rc{
The Anchorage group demonstrated a larger variance in completion time. 
Nine $P_A$ finished the ten tasks faster than average (\ie, $< 27.6$ mins), while only three $P_B$ finished in that time.
From the analytic provenance of three $P_A$ who needed more than 50 minutes, we found that they tended to watch the raw service videos instead of leveraging anchors to prioritize investigative efforts.
This pattern only appeared in one $P_B$.
On the contrary, $P_A$ who finished in 10 minutes were observed focused on validating the anchors and verifying the features.
We expect this would be the norm when the users become familiar with the system.}

\begin{figure}[t]
	\centering
	\includegraphics[width=\linewidth]{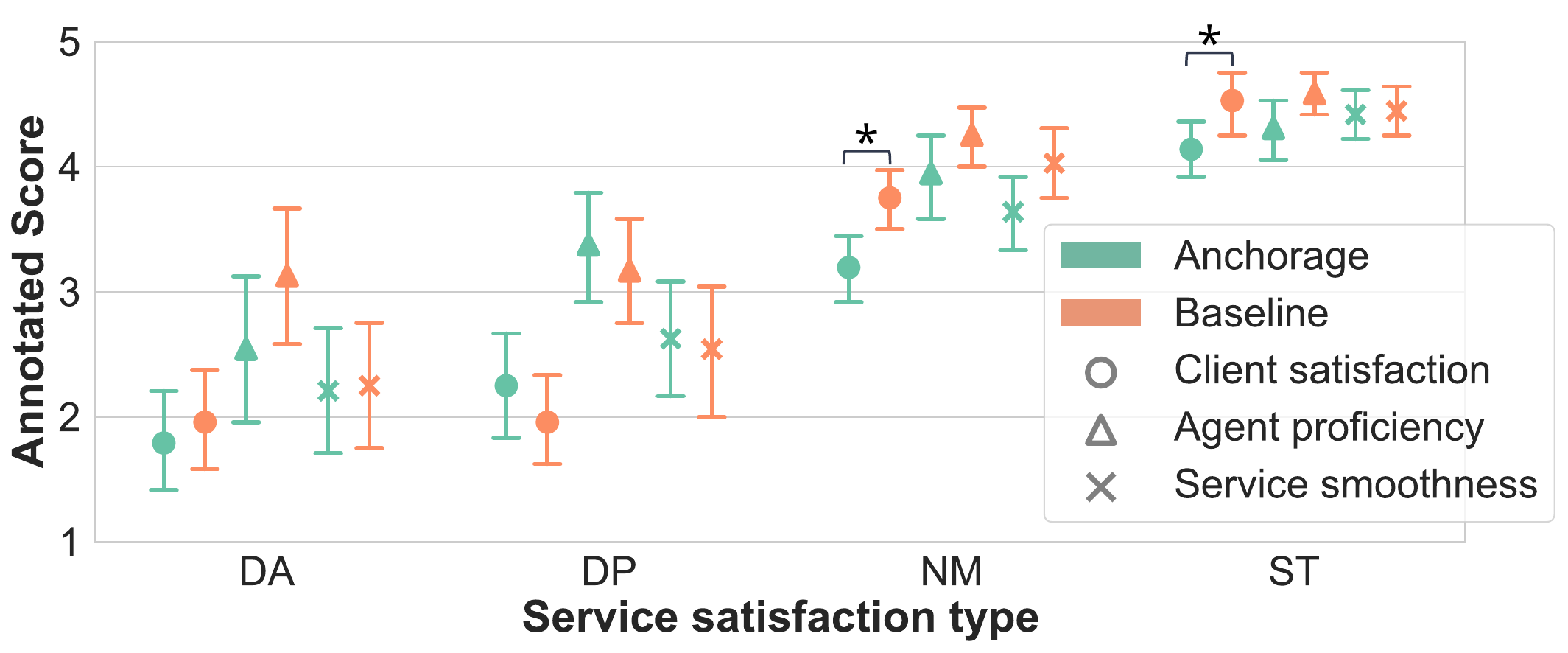}
    \caption{The annotated scores for each video on average and their 95\%CI. \rf{$\ast$ indicates a statistically significant difference.}}
    \label{fig:annotation}
\end{figure}

\textbf{Annotated scores.} 
We compared the three evaluation metrics of each satisfaction type across the Anchorage and Baseline group.
\rf{\autoref{fig:annotation} illustrates the concrete differences in all metrics between satisfaction types.
It indicates that only the difference in client satisfaction score for \textbf{NM} ($U = 32.5$, $p < 0.05$) and \textbf{ST} ($U = 39.5$, $p < 0.05$) between the two groups has statistical significance.}
In general, $P_B$ tended to rate \textbf{NM} and \textbf{ST} with higher satisfaction than $P_A$, whose ratings are closer to the labels provided by \textbf{E5-6}.

We further investigated whether the two systems can clearly distinguish the different satisfaction types.
We performed the Wilcoxon signed-rank tests to accommodate the potential dependencies between tasks.
For both groups, there is no evidence supporting the significance of the agent proficiency between \textbf{NM} and \textbf{ST} ($W_A = 15$, $p_A = 0.202$; $W_B = 5$, $p_B = 0.068$).
All other pair-wise comparisons among satisfaction types in the Anchorage group have $p < 0.05$\rf{, meaning that $P_A$ can clearly distinguish the four satisfaction types}.
However, for the Baseline group, the client satisfaction ($W=10.5$, $p=1$), agent proficiency ($W=18$, $p=1$), and service smoothness ($W=7.5$, $p=0.262$) for the \textbf{DA} and \textbf{DP} pair have no significance.
It suggests that without the support of event contexts in customer services, users might consider the \textbf{DA} and \textbf{DP} types as equals.
This annotation might not be fair to agents who have not caused unsatisfied cases.
Therefore, event contexts should be considered in satisfaction evaluation, even for automatic methods, to prevent biases against agents.

\begin{figure}[t]
	\centering
	\includegraphics[width=\linewidth]{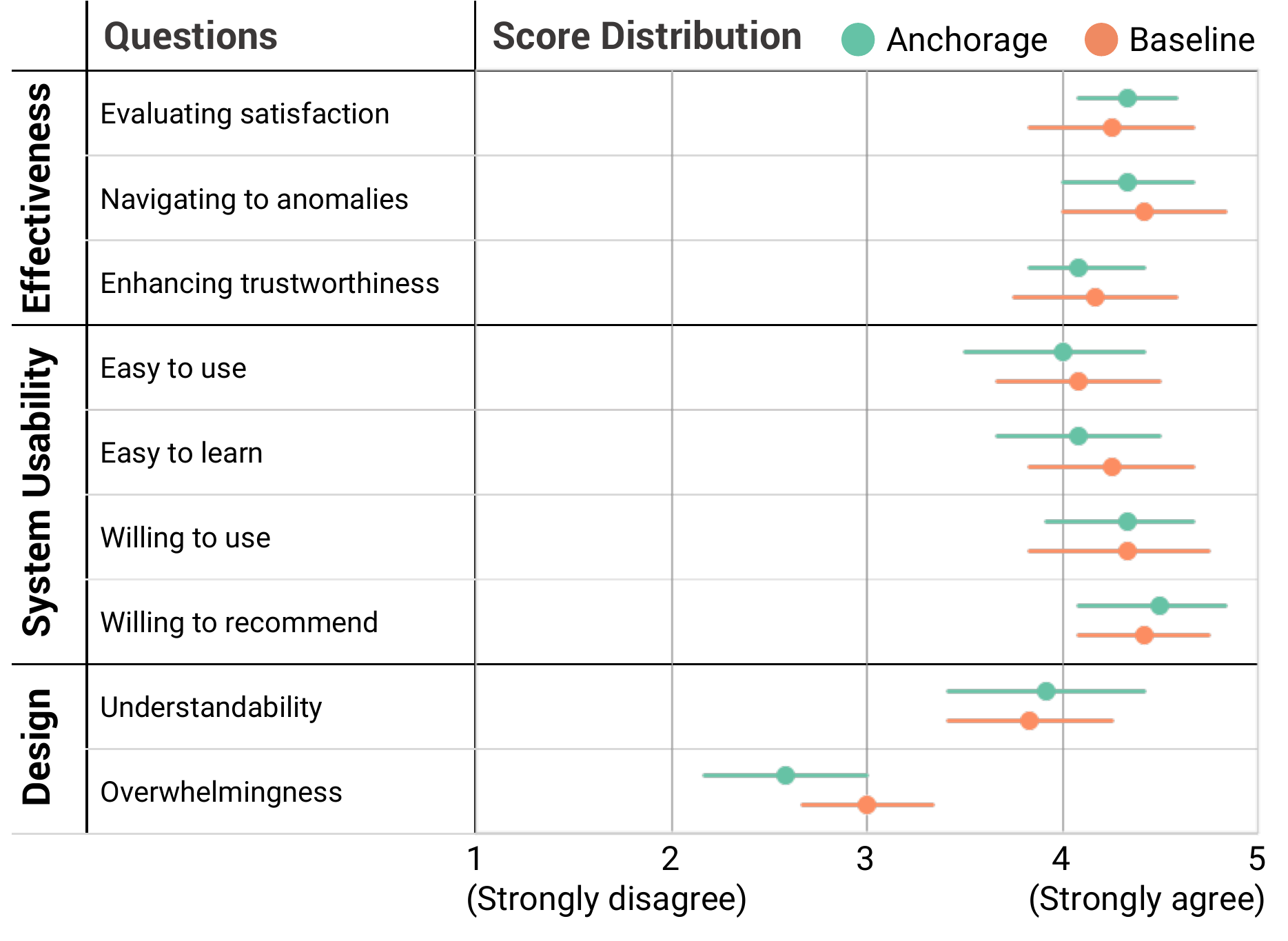}
    \caption{The questionnaire results of the two participant groups in terms of the overall system effectiveness, system usability, and visual designs.}
    \label{fig:user_study}
\end{figure}


\rc{\textbf{Questionnaire.} 
\autoref{fig:user_study} reports the comparable questionnaire results among the two groups in three dimensions: task effectiveness, system usability, and visual designs.
All the metrics listed in \autoref{fig:user_study} have not been found with statistical significance between the two systems.
Using multimodal behavioral features to support satisfaction evaluation tasks are welcomed by practitioners. 
The systems are also able to help quickly navigate anomalies.
Users can distinguish more anomalies (\textbf{DA} vs. \textbf{DP}) using Anchorage with operational anchors than using the baseline system with multimodal behavioral anchors only.
With visual analytics systems, they gained more trust in the automatic results.
Some $P_A$ commented that the scores in individual operations had higher accuracy than the service ones.
This confirms our hypothesis that the introduction of event structure helps evaluate satisfaction.} 

\rc{Participants generally have positive feedback towards the systems for system usability and visual designs.
An interesting finding is that although the baseline system is easier to learn, its understandability and overwhelmingness seem worse than in {\name}.
The slight difference in the learning curve might also be reflected by the slower completion time for T1.
With more visual components and novel visualization designs, the Anchorage is less overwhelming than the Baseline.
It could be attributed to the familiarity of the event contexts and the intuitiveness of the metaphoric designs.
Participants in the Anchorage group have reported more event-related insights, used the quadrant heuristic, and adopted the anchor analogies for the rating rationale.}

\section{Discussion}
\rc{We summarize the lessons learned during the design study about satisfaction analysis in customer service settings as follows.}

\rc{
\textbf{Transferability of customer service.}
We used the public service to characterize the problem domain.
The abstracted workflow is transferable to a customer service dialogue dataset in online shopping~\cite{intro_challenge_chen_2021}, as shown in \autoref{fig:customer-service}.
It is because customer service is characterized by goal-oriented tasks and collaborative communication.
Since our service type focuses on processing applications, our data has more sparse conversations than contact center calls~\cite{related_sat_most_park_2009, related_sat_most_ando_2020}.
However, we also captured dense features such as the clients' facial expressions and designed the system with both types of features. 
It enhances transferability to other service types regardless of feature sparsity.}

\rc{
\textbf{Structuring video with event analysis.}
Videos are often classified as unstructured data because of the difficulties in understanding the states and detecting events.
In situations where the videos are recorded along sequential records, we can adopt event analysis 
to formulate granular video segmentation schemes.
These scenarios will become more prevalent as remote work environments become increasingly popular. 
The framework described in this work can be generalized to other applications, such as online education, smart manufacturing, system interface testing, and interactions in virtual reality.
We showed that introducing operational anchors into the conventional video-based satisfaction analysis streamlines video content understanding.
Video analytics should eagerly look for event structures to frame the features extracted from videos. }

\rc{
\textbf{Improvised dataset as test cases for VAST.}
Collecting a real dataset is one of the most difficult challenges for VAST systems~\cite{background_vis_sedlmair_2012}, including video analytics. 
In satisfaction analysis, extreme cases (very satisfied and very dissatisfied) are rare in real life.
We used the improvised dataset to tackle the sample imbalance challenge.
With close collaboration with domain experts, we define the typical satisfaction patterns and collect these patterns through improvising.
They provide ground truth that can be used for pretraining models and act as the initial reference points to mitigate the cold-start problems.
We can also control the environment to isolate unwanted effects and make the results comparable.}

\rc{
Incorporating domain experts to participate in creating such a dataset has been beneficial to our study. 
The dataset laid the common ground for our collaborators and us to discuss the expectation over the visual designs during the design process.
However, in our case, the authenticity of the client's response and dynamics is disputable due to improvision.
For example, customers may look at their phones more often when they feel bored by a long process~\cite{related_sat_app_liu_2019}, while we asked our subjects to be more attentive.
We should also be mindful of the Observer's Paradox, which states that experimenters' presence influences data gathering~\cite{related_sat_video_gunes_2013, related_sat_video_mcduff_2012}. 
More standardized protocols should be discussed and developed to promote fair and just evaluation.}

\textbf{Privacy and multimodal features.}
Recording the services is a norm in customer services to prevent conflicts resulting from misinformation~\cite{intro_service_saberi_2017}.
\rf{Videos} are particularly useful when complaints need to be investigated.
\rf{In this work, the customer service videos were collected with internal IRB's approval (\#HREP-2021-0162) and written consent from participants.
However, the increasingly tightening privacy policies might forbid the collection of service videos that include sensitive features such as the clients' faces.
Since {\name} utilizes many emotional features extracted from the videos, its effectiveness would be significantly affected if the video collections were prohibited.}

However, we have shown that combining operational and behavioral anchors can enhance satisfaction evaluation performance \rf{without compromising annotation precision}.
There are more multimodal features worth exploring that could facilitate satisfaction evaluation.
\rf{For example, event features (\eg, business procedures) and other behavioral features (\eg, machine operations and agent behaviors) provide contexts for causal analysis to satisfaction.
Screen-space and data-space sanitization techniques~\cite{discussion_privacy_zhou_2023} could be useful in protecting clients' privacy.
We have also collected multi-view videos about the clients which can provide additional viewpoints to infer their actions.
Future works can explore visualization techniques for summarizing multi-view videos.
Another under-explored feature is the agents' emotions. 
The emotional interaction between the agent and client could infer the other party's affective status and evaluate whether the agent reacts appropriately.}
\section{Conclusion}
This paper investigates customer satisfaction evaluation with customer service videos and service records.
The fusion of multimodal behavioral features extracted from videos provides a primitive satisfaction evaluation.
We introduce the use of machine logs to provide semantically meaningful video understanding and magnify a conventional satisfaction score with greater sequential and temporal resolutions.
They both constitute the anchor concept.
We constructed the anchors with a multi-perspective anomaly detection framework to help narrow down the vast event space.
We developed the buoy charts and multi-faceted views to effectively summarize the services and navigate users to segments of interest.

We created an improvised dataset to show that {\name} can detect satisfactory, unsatisfactory, and counteract normal cases.
The combination of video analytics and event sequence analysis shows promising results in effectively understanding videos.
We found that introducing event contexts to video analytics can enhance the performance of evaluating customer satisfaction in videos.
Our approach can be adapted in situations where unlabelled and unstructured videos are collected along with sequential records.

\section*{Acknowledgments}
We are grateful to Wucong Chen and the Guang Fo Zhan Qu team for their helpful feedback and coordination for the user studies. We also thank anonymous reviewers for their constructive comments. This research was supported in part by grant FSUST19-CWB09. 



\bibliographystyle{IEEEtran}
\bibliography{main}

\begin{thebibliography}{10}
\providecommand{\url}[1]{#1}
\csname url@samestyle\endcsname
\providecommand{\newblock}{\relax}
\providecommand{\bibinfo}[2]{#2}
\providecommand{\BIBentrySTDinterwordspacing}{\spaceskip=0pt\relax}
\providecommand{\BIBentryALTinterwordstretchfactor}{4}
\providecommand{\BIBentryALTinterwordspacing}{\spaceskip=\fontdimen2\font plus
\BIBentryALTinterwordstretchfactor\fontdimen3\font minus
  \fontdimen4\font\relax}
\providecommand{\BIBforeignlanguage}[2]{{%
\expandafter\ifx\csname l@#1\endcsname\relax
\typeout{** WARNING: IEEEtran.bst: No hyphenation pattern has been}%
\typeout{** loaded for the language `#1'. Using the pattern for}%
\typeout{** the default language instead.}%
\else
\language=\csname l@#1\endcsname
\fi
#2}}
\providecommand{\BIBdecl}{\relax}
\BIBdecl

\bibitem{background_satisfaction_oliver_2014}
R.~L. Oliver, \emph{Satisfaction: A behavioral perspective on the consumer},
  2nd~ed.\hskip 1em plus 0.5em minus 0.4em\relax Routledge, 2010.

\bibitem{intro_motivation_iso_2018}
``\textit{Quality management — Customer satisfaction — Guidelines for
  monitoring and measuring},'' ISO 10004:2018, 2018.

\bibitem{intro_emotion_wong_2004}
A.~Wong, ``The role of emotional satisfaction in service encounters,''
  \emph{Managing Service Quality: An International Journal}, vol.~14, no.~5,
  pp. 365--376, 2004.

\bibitem{intro_motivation_tronvoll_2011}
B.~Tronvoll, ``Negative emotions and their effect on customer complaint
  behaviour,'' \emph{Journal of Service Management}, vol.~22, no.~1, pp.
  111--134, 2011.

\bibitem{intro_motivation_peterson_1992}
R.~A. Peterson and W.~R. Wilson, ``Measuring customer satisfaction: fact and
  artifact,'' \emph{Journal of the academy of marketing science}, vol.~20,
  no.~1, pp. 61--71, 1992.

\bibitem{related_sat_app_alotaibi_2018}
S.~Al-Otaibi, A.~Alnassar, A.~Alshahrani, A.~Al-Mubarak, S.~Albugami,
  N.~Almutiri, and A.~Albugami, ``Customer satisfaction measurement using
  sentiment analysis,'' \emph{International Journal of Advanced Computer
  Science and Applications}, vol.~9, no.~2, pp. 106--117, 2018.

\bibitem{related_sat_app_see_2021}
A.~See and C.~Manning, ``Understanding and predicting user dissatisfaction in a
  neural generative chatbot,'' in \emph{Proc. SIGDIAL}.\hskip 1em plus 0.5em
  minus 0.4em\relax Singapore and Online: Association for Computational
  Linguistics, Jul. 2021, pp. 1--12.

\bibitem{related_sat_most_park_2009}
Y.~Park and S.~C. Gates, ``Towards real-time measurement of customer
  satisfaction using automatically generated call transcripts,'' in \emph{Proc.
  CIKM}.\hskip 1em plus 0.5em minus 0.4em\relax NY: ACM, 2009, p. 1387–1396.

\bibitem{related_sat_most_ando_2020}
A.~Ando, R.~Masumura, H.~Kamiyama, S.~Kobashikawa, Y.~Aono, and T.~Toda,
  ``Customer satisfaction estimation in contact center calls based on a
  hierarchical multi-task model,'' \emph{IEEE/ACM Transactions on Audio,
  Speech, and Language Processing}, vol.~28, pp. 715--728, 2020.

\bibitem{intro_service_saberi_2017}
M.~Saberi, O.~K. Hussain, and E.~Chang, ``Past, present and future of contact
  centers: a literature review,'' \emph{Business Process Management Journal},
  2017.

\bibitem{related_sat_most_seng_2018}
K.~P. Seng and L.-M. Ang, ``Video analytics for customer emotion and
  satisfaction at contact centers,'' \emph{IEEE Transactions on Human-Machine
  Systems}, vol.~48, no.~3, pp. 266--278, 2018.

\bibitem{intro_emotion_liljander_1997}
V.~Liljander and T.~Strandvik, ``Emotions in service satisfaction,''
  \emph{International Journal of Service Industry Management}, vol.~8, no.~2,
  pp. 148--169, 1997.

\bibitem{background_satisfaction_cheshin_2018}
A.~Cheshin, A.~Amit, and G.~A. {van Kleef}, ``The interpersonal effects of
  emotion intensity in customer service: Perceived appropriateness and
  authenticity of attendants' emotional displays shape customer trust and
  satisfaction,'' \emph{Organizational Behavior and Human Decision Processes},
  vol. 144, pp. 97--111, 2018.

\bibitem{related_sat_app_zhang_2020}
Q.~Zhang, W.~Wang, and Y.~Chen, ``Frontiers: In-consumption social listening
  with moment-to-moment unstructured data: The case of movie appreciation and
  live comments,'' \emph{Marketing Science}, vol.~39, no.~2, pp. 285--295,
  2020.

\bibitem{related_sat_app_liu_2019}
Z.~X. Liu, Y.~Liu, and X.~Gao, ``Using mobile eye tracking to evaluate the
  satisfaction with service office,'' in \emph{Design, User Experience, and
  Usability. Practice and Case Studies}.\hskip 1em plus 0.5em minus 0.4em\relax
  Cham: Springer International Publishing, 2019, pp. 183--195.

\bibitem{related_sat_app_kumar_2019}
S.~Kumar, M.~Yadava, and P.~P. Roy, ``Fusion of eeg response and sentiment
  analysis of products review to predict customer satisfaction,''
  \emph{Information Fusion}, vol.~52, pp. 41--52, 2019.

\bibitem{related_sat_video_gunes_2013}
H.~Gunes and B.~Schuller, ``Categorical and dimensional affect analysis in
  continuous input: Current trends and future directions,'' \emph{Image and
  Vision Computing}, vol.~31, no.~2, pp. 120--136, 2013.

\bibitem{related_sat_video_mcduff_2012}
D.~McDuff, R.~E. Kaliouby, and R.~W. Picard, ``Crowdsourcing facial responses
  to online videos,'' \emph{IEEE Transactions on Affective Computing}, vol.~3,
  no.~4, pp. 456--468, 2012.

\bibitem{related_sat_video_generosi_2018}
A.~Generosi, S.~Ceccacci, and M.~Mengoni, ``A deep learning-based system to
  track and analyze customer behavior in retail store,'' in \emph{Proc.
  ICCE-Berlin}, 2018, pp. 1--6.

\bibitem{related_sat_video_mcduff_2015}
D.~McDuff, R.~E. Kaliouby, J.~F. Cohn, and R.~W. Picard, ``Predicting ad liking
  and purchase intent: Large-scale analysis of facial responses to ads,''
  \emph{IEEE Transactions on Affective Computing}, vol.~6, no.~3, pp. 223--235,
  2015.

\bibitem{related_sat_video_slim_2018}
M.~Slim, R.~Kachouri, and A.~B. Atitallah, ``Customer satisfaction measuring
  based on the most significant facial emotion,'' in \emph{Proc. SSD}, 2018,
  pp. 502--507.

\bibitem{related_sat_video_sugianto_2018}
N.~Sugianto, D.~Tjondronegoro, and B.~Tydd, ``Deep residual learning for
  analyzing customer satisfaction using video surveillance,'' in \emph{Proc.
  AVSS}, 2018, pp. 1--6.

\bibitem{related_sat_video_gonzalez_2020}
M.~González-Rodríguez, M.~Díaz-Fernández, and C.~{Pacheco Gómez},
  ``Facial-expression recognition: An emergent approach to the measurement of
  tourist satisfaction through emotions,'' \emph{Telematics and Informatics},
  vol.~51, p. 101404, 2020.

\bibitem{related_sat_video_yolcu_2018}
G.~Yolcu, I.~Oztel, S.~Kazan, C.~Oz, and F.~Bunyak, ``Deep learning-based face
  analysis system for monitoring customer interest,'' \emph{Journal of ambient
  intelligence and humanized computing}, vol.~11, no.~1, pp. 237--248, 2020.

\bibitem{related_video_emotion_zeng_2020}
H.~Zeng, X.~Wang, A.~Wu, Y.~Wang, Q.~Li, A.~Endert, and H.~Qu, ``{EmoCo}:
  Visual analysis of emotion coherence in presentation videos,'' \emph{IEEE
  Transactions on Visualization and Computer Graphics}, vol.~26, no.~1, pp.
  927--937, 2020.

\bibitem{related_video_emotion_ma_2020}
C.-X. Ma, J.-C. Song, Q.~Zhu, K.~Maher, Z.-Y. Huang, and H.-A. Wang,
  ``Emotionmap: Visual analysis of video emotional content on a map,''
  \emph{Journal of Computer Science and Technology}, vol.~35, no.~3, pp.
  576--591, 2020.

\bibitem{related_video_voicecoach_wang_2020}
X.~Wang, H.~Zeng, Y.~Wang, A.~Wu, Z.~Sun, X.~Ma, and H.~Qu, ``Voicecoach:
  Interactive evidence-based training for voice modulation skills in public
  speaking,'' in \emph{Proc. CHI}.\hskip 1em plus 0.5em minus 0.4em\relax New
  York, NY, USA: ACM, 2020, p. 1–12.

\bibitem{related_video_dehumor_wang_2021}
X.~Wang, Y.~Ming, T.~Wu, H.~Zeng, Y.~Wang, and H.~Qu, ``Dehumor: Visual
  analytics for decomposing humor,'' \emph{IEEE Transactions on Visualization
  and Computer Graphics}, vol.~28, no.~12, pp. 4609--4623, 2022.

\bibitem{related_video_emotion_maher_2022}
K.~Maher, Z.~Huang, J.~Song, X.~Deng, Y.-K. Lai, C.~Ma, H.~Wang, Y.-J. Liu, and
  H.~Wang, ``E-ffective: A visual analytic system for exploring the emotion and
  effectiveness of inspirational speeches,'' \emph{IEEE Transactions on
  Visualization and Computer Graphics}, vol.~28, no.~1, pp. 508--517, 2022.

\bibitem{related_video_gesturelens_2022}
H.~Zeng, X.~Wang, Y.~Wang, A.~Wu, T.-C. Pong, and H.~Qu, ``Gesturelens: Visual
  analysis of gestures in presentation videos,'' \emph{IEEE Transactions on
  Visualization and Computer Graphics}, pp. 1--1, 2022.

\bibitem{related_video_multimodal_zeng_2021}
H.~Zeng, X.~Shu, Y.~Wang, Y.~Wang, L.~Zhang, T.-C. Pong, and H.~Qu,
  ``{E}motion{C}ues: Emotion-oriented visual summarization of classroom
  videos,'' \emph{IEEE Transactions on Visualization and Computer Graphics},
  vol.~27, no.~7, pp. 3168--3181, 2021.

\bibitem{related_video_multimodal_li_2021}
H.~Li, M.~Xu, Y.~Wang, H.~Wei, and H.~Qu, ``A visual analytics approach to
  facilitate the proctoring of online exams,'' in \emph{Proc. CHI}.\hskip 1em
  plus 0.5em minus 0.4em\relax NY: ACM, 2021.

\bibitem{related_video_multimodal_tang_2022}
T.~Tang, Y.~Wu, Y.~Wu, L.~Yu, and Y.~Li, ``Videomoderator: A risk-aware
  framework for multimodal video moderation in e-commerce,'' \emph{IEEE
  Transactions on Visualization and Computer Graphics}, vol.~28, no.~1, pp.
  846--856, 2022.

\bibitem{related_video_multiviz_2022}
P.~P. Liang, Y.~Lyu, G.~Chhablani, N.~Jain, Z.~Deng, X.~Wang, L.-P. Morency,
  and R.~Salakhutdinov, ``Multiviz: An analysis benchmark for visualizing and
  understanding multimodal models,'' \emph{ArXiv preprint ArXiv:2207.00056},
  2022.

\bibitem{related_video_multimodal_wang_2022}
X.~Wang, J.~He, Z.~Jin, M.~Yang, Y.~Wang, and H.~Qu, ``{M2L}ens: Visualizing
  and explaining multimodal models for sentiment analysis,'' \emph{IEEE
  Transactions on Visualization and Computer Graphics}, vol.~28, no.~1, pp.
  802--812, 2022.

\bibitem{related_video_multimodal_wu_2020}
A.~Wu and H.~Qu, ``Multimodal analysis of video collections: Visual exploration
  of presentation techniques in ted talks,'' \emph{IEEE Transactions on
  Visualization and Computer Graphics}, vol.~26, no.~7, pp. 2429--2442, 2020.

\bibitem{related_video_multimodal_blascheck_2016}
T.~Blascheck, F.~Beck, S.~Baltes, T.~Ertl, and D.~Weiskopf, ``Visual analysis
  and coding of data-rich user behavior,'' in \emph{Proc. VAST}, 2016, pp.
  141--150.

\bibitem{related_video_multimodal_soure_2022}
E.~J. Soure, E.~Kuang, M.~Fan, and J.~Zhao, ``Coux: Collaborative visual
  analysis of think-aloud usability test videos for digital interfaces,''
  \emph{IEEE Transactions on Visualization and Computer Graphics}, vol.~28,
  no.~1, pp. 643--653, 2022.

\bibitem{discussion_dataset_lin_2021}
Y.~Lin, K.~Wong, Y.~Wang, R.~Zhang, B.~Dong, H.~Qu, and Q.~Zheng, ``Taxthemis:
  Interactive mining and exploration of suspicious tax evasion groups,''
  \emph{IEEE Transactions on Visualization and Computer Graphics}, vol.~27,
  no.~2, pp. 849--859, 2021.

\bibitem{discussion_dataset_zhang_2023}
W.~Zhang, J.~K. Wong, X.~Wang, Y.~Gong, R.~Zhu, K.~Liu, Z.~Yan, S.~Tan, H.~Qu,
  S.~Chen, and W.~Chen, ``Cohortva: A visual analytic system for interactive
  exploration of cohorts based on historical data,'' \emph{IEEE Transactions on
  Visualization and Computer Graphics}, vol.~29, no.~1, pp. 756--766, 2023.

\bibitem{related_video_survey_2015}
B.~H\"{o}ferlin, M.~H\"{o}ferlin, G.~Heidemann, and D.~Weiskopf, ``Scalable
  video visual analytics,'' \emph{Information Visualization}, vol.~14, no.~1,
  pp. 10--26, 2015.

\bibitem{related_video_event_kurzhals_2016}
K.~Kurzhals, M.~John, F.~Heimerl, P.~Kuznecov, and D.~Weiskopf, ``Visual movie
  analytics,'' \emph{IEEE Transactions on Multimedia}, vol.~18, no.~11, pp.
  2149--2160, 2016.

\bibitem{related_video_event_chen_2022}
Z.~Chen, S.~Ye, X.~Chu, H.~Xia, H.~Zhang, H.~Qu, and Y.~Wu, ``Augmenting sports
  videos with viscommentator,'' \emph{IEEE Transactions on Visualization and
  Computer Graphics}, vol.~28, no.~1, pp. 824--834, 2022.

\bibitem{related_video_event_deng_2021}
D.~Deng, J.~Wu, J.~Wang, Y.~Wu, X.~Xie, Z.~Zhou, H.~Zhang, X.~L. Zhang, and
  Y.~Wu, ``Eventanchor: Reducing human interactions in event annotation of
  racket sports videos,'' in \emph{Proc. CHI}.\hskip 1em plus 0.5em minus
  0.4em\relax NY: ACM, 2021.

\bibitem{related_event_app_shi_2014}
C.~Shi, Y.~Wu, S.~Liu, H.~Zhou, and H.~Qu, ``Loyaltracker: Visualizing loyalty
  dynamics in search engines,'' \emph{IEEE Transactions on Visualization and
  Computer Graphics}, vol.~20, no.~12, pp. 1733--1742, 2014.

\bibitem{related_event_hierarchy_cappers_2018}
B.~C. Cappers and J.~J. van Wijk, ``Exploring multivariate event sequences
  using rules, aggregations, and selections,'' \emph{IEEE Transactions on
  Visualization and Computer Graphics}, vol.~24, no.~1, pp. 532--541, 2018.

\bibitem{related_event_hierarchy_magallanes_2022}
J.~Magallanes, T.~Stone, P.~D. Morris, S.~Mason, S.~Wood, and M.-C.
  Villa-Uriol, ``Sequen-c: A multilevel overview of temporal event sequences,''
  \emph{IEEE Transactions on Visualization and Computer Graphics}, vol.~28,
  no.~1, pp. 901--911, 2022.

\bibitem{related_event_hierarchy_liu_2017}
Z.~Liu, B.~Kerr, M.~Dontcheva, J.~Grover, M.~Hoffman, and A.~Wilson,
  ``Coreflow: Extracting and visualizing branching patterns from event
  sequences,'' \emph{Computer Graphics Forum}, vol.~36, no.~3, pp. 527--538,
  2017.

\bibitem{related_event_hierarchy_polack_2018}
P.~J. Polack~Jr., S.-T. Chen, M.~Kahng, K.~D. Barbaro, R.~Basole, M.~Sharmin,
  and D.~H. Chau, ``Chronodes: Interactive multifocus exploration of event
  sequences,'' \emph{ACM Trans. Interact. Intell. Syst.}, vol.~8, no.~1, feb
  2018.

\bibitem{related_event_survey_guo_2021}
Y.~Guo, S.~Guo, Z.~Jin, S.~Kaul, D.~Gotz, and N.~Cao, ``A survey on visual
  analysis of event sequence data,'' \emph{IEEE Transactions on Visualization
  and Computer Graphics}, pp. 1--1, 2021.

\bibitem{related_event_anomaly_nguyen_2019}
P.~H. Nguyen, C.~Turkay, G.~Andrienko, N.~Andrienko, O.~Thonnard, and
  J.~Zouaoui, ``Understanding user behaviour through action sequences: From the
  usual to the unusual,'' \emph{IEEE Transactions on Visualization and Computer
  Graphics}, vol.~25, no.~9, pp. 2838--2852, 2019.

\bibitem{related_event_anomaly_yeshchenko_2021}
A.~Yeshchenko, C.~Di~Ciccio, J.~Mendling, and A.~Polyvyanyy, ``Visual drift
  detection for sequence data analysis of business processes,'' \emph{IEEE
  Transactions on Visualization and Computer Graphics}, pp. 1--1, 2021.

\bibitem{related_event_anomaly_guo_2021}
S.~Guo, Z.~Jin, Q.~Chen, D.~Gotz, H.~Zha, and N.~Cao, ``Interpretable anomaly
  detection in event sequences via sequence matching and visual comparison,''
  \emph{IEEE Transactions on Visualization and Computer Graphics}, pp. 1--1,
  2021.

\bibitem{intro_challenge_chen_2021}
D.~Chen, H.~Chen, Y.~Yang, A.~Lin, and Z.~Yu, ``Action-based conversations
  dataset: A corpus for building more in-depth task-oriented dialogue
  systems,'' in \emph{Proc. {NAACL-HLT}}.\hskip 1em plus 0.5em minus
  0.4em\relax Online: ACL, Jun. 2021, pp. 3002--3017.

\bibitem{background_vis_sedlmair_2012}
M.~Sedlmair, M.~Meyer, and T.~Munzner, ``Design study methodology: Reflections
  from the trenches and the stacks,'' \emph{IEEE Transactions on Visualization
  and Computer Graphics}, vol.~18, no.~12, pp. 2431--2440, 2012.

\bibitem{method_visual_qi_2021}
D.~Qi, W.~Tan, Q.~Yao, and J.~Liu, ``Yolo5face: Why reinventing a face
  detector,'' \emph{ArXiv preprint ArXiv:2105.12931}, 2021.

\bibitem{method_visual_li_2020}
S.~Li and W.~Deng, ``Deep facial expression recognition: A survey,'' \emph{IEEE
  Transactions on Affective Computing}, pp. 1--1, 2020.

\bibitem{method_visual_yang_2019}
T.-Y. Yang, Y.-T. Chen, Y.-Y. Lin, and Y.-Y. Chuang, ``Fsa-net: Learning
  fine-grained structure aggregation for head pose estimation from a single
  image,'' in \emph{Proc. CVPR}, 2019, pp. 1087--1096.

\bibitem{method_visual_wang_2020}
K.~Wang, X.~Peng, J.~Yang, S.~Lu, and Y.~Qiao, ``Suppressing uncertainties for
  large-scale facial expression recognition,'' in \emph{Proc. CVPR}, June 2020.

\bibitem{method_audio_bredin_2020}
H.~Bredin, R.~Yin, J.~M. Coria, G.~Gelly, P.~Korshunov, M.~Lavechin, D.~Fustes,
  H.~Titeux, W.~Bouaziz, and M.-P. Gill, ``Pyannote.audio: Neural building
  blocks for speaker diarization,'' in \emph{IEEE International Conference on
  Acoustics, Speech, and Signal Processing}, 2020, pp. 7124--7128.

\bibitem{method_audio_pinto_2021}
M.~G. de~Pinto, M.~Polignano, P.~Lops, and G.~Semeraro, ``Emotions
  understanding model from spoken language using deep neural networks and
  mel-frequency cepstral coefficients,'' in \emph{2020 IEEE Conference on
  Evolving and Adaptive Intelligent Systems (EAIS)}, 2020, pp. 1--5.

\bibitem{method_anchor_xu_2009}
W.~Xu, L.~Huang, A.~Fox, D.~Patterson, and M.~I. Jordan, ``Detecting
  large-scale system problems by mining console logs,'' in \emph{Proc.
  SIGOPS}.\hskip 1em plus 0.5em minus 0.4em\relax New York, NY, USA: ACM, 2009,
  p. 117–132.

\bibitem{method_anchor_lou_2010}
J.-G. Lou, Q.~Fu, S.~Yang, Y.~Xu, and J.~Li, ``Mining invariants from console
  logs for system problem detection,'' in \emph{Proc. USENIX}, ser.
  USENIXATC'10.\hskip 1em plus 0.5em minus 0.4em\relax USA: USENIX Association,
  2010, p.~24.

\bibitem{method_anchor_ye_2000}
N.~Ye, ``A markov chain model of temporal behavior for anomaly detection,'' in
  \emph{Proceedings of the IEEE Systems, Man, and Cybernetics Information
  Assurance and Security Workshop}, vol. 166, 2000, p. 169.

\bibitem{visual_design_yang_2021}
L.~Yang, C.~Xiong, J.~K. Wong, A.~Wu, and H.~Qu, ``Explaining with examples
  lessons learned from crowdsourced introductory description of information
  visualizations,'' \emph{IEEE Transactions on Visualization and Computer
  Graphics}, pp. 1--1, 2021.

\bibitem{visual_design_gleicher_2018}
M.~Gleicher, ``Considerations for visualizing comparison,'' \emph{IEEE
  Transactions on Visualization and Computer Graphics}, vol.~24, no.~1, pp.
  413--423, 2018.

\bibitem{visual_multimodal_morrow_2019}
B.~Morrow, T.~Manz, A.~E. Chung, N.~Gehlenborg, and D.~Gotz, ``Periphery plots
  for contextualizing heterogeneous time-based charts,'' in \emph{2019 IEEE
  Visualization Conference (VIS)}, 2019, pp. 1--5.

\bibitem{discussion_privacy_zhou_2023}
J.~Zhou, X.~Wang, J.~K. Wong, H.~Wang, Z.~Wang, X.~Yang, X.~Yan, H.~Feng,
  H.~Qu, H.~Ying, and W.~Chen, ``Dpviscreator: Incorporating pattern
  constraints to privacy-preserving visualizations via differential privacy,''
  \emph{IEEE Transactions on Visualization and Computer Graphics}, vol.~29,
  no.~1, pp. 809--819, 2023.

\end{thebibliography}

\begin{IEEEbiography}[{\includegraphics[width=1in,height=1.25in,clip,keepaspectratio]{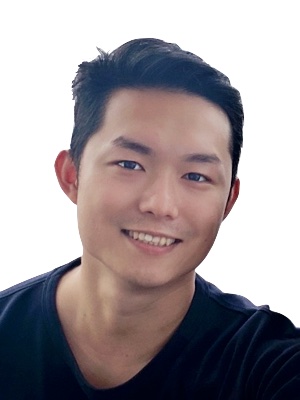}}]{Kam Kwai Wong} 
is a Ph.D. candidate in the Department of Computer Science and Engineering at the Hong Kong University of Science and Technology (HKUST). He received his B.E. in HKUST. His main research interests are in data visualization, visual analytics and data mining.
For more details, please refer to \url{https://jasonwong.vision}.
\end{IEEEbiography}

\begin{IEEEbiography}[{\includegraphics[width=1in,height=1.25in,clip,keepaspectratio]{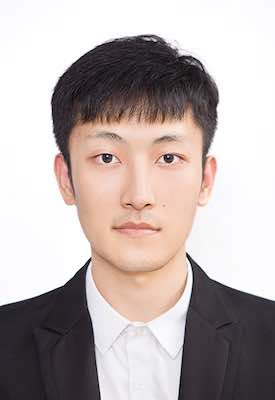}}]{Xingbo Wang}
is a Ph.D. candidate in the Department of Computer Science and Engineering at the Hong Kong University of Science and Technology (HKUST). He obtained a B.E. degree from Wuhan University, China in 2018. His research interests include human-computer interaction (HCI), data visualization, natural language processing (NLP), and multimodal analysis. For more details, please refer to \url{https://andy-xingbowang.com}.
\end{IEEEbiography}

\begin{IEEEbiography}[{\includegraphics[width=1in,height=1.25in,clip,keepaspectratio]{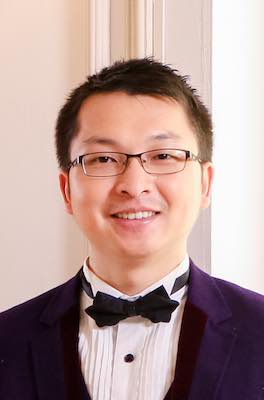}}]{Yong Wang} is an assistant professor in School of Computing and Information Systems at Singapore Management University. His research interests include data visualization, visual analytics and explainable machine learning.
He obtained his Ph.D. in Computer Science from Hong Kong University of Science and Technology in 2018. He received his B.E. and M.E. from Harbin Institute of Technology and Huazhong University of Science and Technology, respectively. For more details, please refer to \url{http://yong-wang.org}.
\end{IEEEbiography}

\begin{IEEEbiography}[{\includegraphics[width=1in,height=1.25in,clip,keepaspectratio]{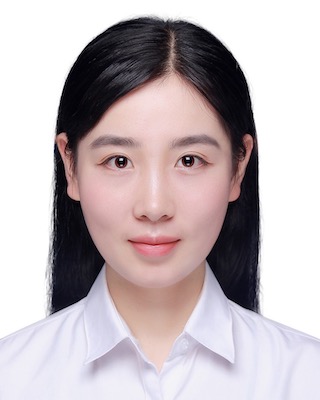}}]{Jianben He} 
is a Ph.D. candidate in the Department of Computer Science and Engineering at the Hong Kong University of Science and Technology (HKUST). She obtained a B.E. degree in Electronic Information and Communication from Huazhong University of Science and Technology. For more details, please refer to \url{https://jessiehe970311.github.io/}.
\end{IEEEbiography}

\begin{IEEEbiography}[{\includegraphics[width=1in,height=1.25in,clip,keepaspectratio]{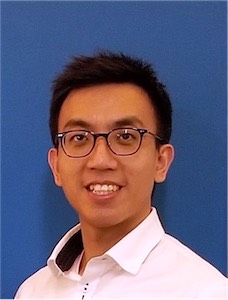}}]{Rong Zhang}
is the Lecturer in the Interdisciplinary Programs Office of HKUST. His research interests include data visualization, human-centered design, and interdisciplinary education.
\end{IEEEbiography}

\begin{IEEEbiography}[{\includegraphics[width=1in,height=1.25in,clip,keepaspectratio]{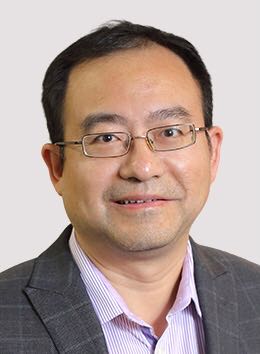}}]{Huamin Qu} 
is a chair professor in the Department of Computer Science and Engineering (CSE), the director of the Interdisciplinary Program Office (IPO), and the head of the Division of Emerging Interdisciplinary Areas (EMIA) at the Hong Kong University of Science and Technology (HKUST). He obtained a B.S. in Mathematics from Xi'an Jiaotong University, China, an M.S. and a Ph.D. in Computer Science from the Stony Brook University. He is the director of the VisLab and also serves as the coordinator of the Human-Computer Interaction (HCI) group at the CSE department. His main research interests are in visualization and human-computer interaction, with focuses on urban informatics, social network analysis, E-learning, text visualization, and explainable artificial intelligence (XAI).
\end{IEEEbiography}



\end{document}